\documentclass[preprint,12pt]{elsarticle}

\usepackage{graphicx}
\usepackage{amssymb}
\usepackage{amsmath}
\usepackage{xcolor}
\usepackage{lineno}

\journal{Journal Name}

\begin{document}

\begin{frontmatter}


\title{\textcolor{black}{Multiscale method based on coupled lattice-Boltzmann and Langevin-dynamics for direct simulation of nanoscale particle/polymer suspensions in complex flows}}



\author[1,2]{Zixiang Liu}
\author[1]{Yuanzheng Zhu}
\author[3]{Jonathan R. Clausen}
\author[3]{Jeremy B. Lechman}
\author[3]{Rekha R. Rao}
\author[1,2]{Cyrus K. Aidun}

\address[1]{George W. Woodruff School of Mechanical Engineering, Georgia Institute of Technology, Atlanta, GA, 30332, USA}
\address[2]{Parker H. Petit Institute for Bioengineering and Bioscience, Georgia Institute of Technology, Atlanta, GA, 30322, USA}
\address[3]{Sandia National Laboratories, Albuquerque, NM, 87185, USA}

\begin{abstract}
A hybrid computational method coupling the lattice-Boltzmann (LB) method and a Langevin-dynamics (LD) method is developed to simulate nanoscale particle and polymer (NPP) suspensions in the presence of both thermal fluctuation and long-range many-body hydrodynamic interactions (HI). Brownian motion of the NPP is explicitly captured by a stochastic forcing term in the LD method. The LD method is two-way coupled to the non-fluctuating LB fluid through a discrete LB forcing source distribution to capture the long-range HI. To ensure intrinsically linear scalability with respect to the number of particles, an Eulerian-host algorithm for short-distance particle neighbor search and interaction is developed and embedded to LB-LD framework. The validity and accuracy of the LB-LD approach are demonstrated through several sample problems. The simulation results show good agreements with theory and experiment. The LB-LD approach can be favorably incorporated into complex multiscale computational frameworks for efficiently simulating multiscale, multicomponent particulate suspension systems such as complex blood suspensions.
\end{abstract}


\end{frontmatter}



\section{Introduction}\label{sec:intro}
Simulating suspensions of nanoscale particles or polymers is a challenging task that involves resolving particle-particle interaction, thermal fluctuation and long-range many-body hydrodynamic interactions (HI) that contribute to complex suspension microstructures \citep{russel1981,ladd2001}. Conventional particle-based methods such as Molecular dynamics (MD) \citep{Dunweg1993}, dissipative particle dynamics (DPD) \citep{groot1997}, Brownian dynamics (BD) \citep{mcc1978} and Stokesian dynamics (SD) \citep{Brady1988} have advanced this field considerably. The MD and DPD approaches resolve the fluid particles explicitly and are often suitable for problem with relatively small length and time scales. The conventional BD neglects the particle inertia the HI effect to accommodate for computational efficiency \citep{mcc1978}. The SD method has been a successful computational tool to study the rheology of spherical, rigid particle suspensions. It captures the many-body HI theoretically with excellent rheological agreement with experiments \citep{FossJFM1999,FossJFM2000,Sierou2004}. Although it intrinsically scale cubically with the number of particles, $N$, recent attempts have been made to enhance SD to O(NlnN) or O(N) scales through sophisticated computational algorithms \citep{Brady2003,sierou2001,graham2012,wang2016}. Nevertheless, the SD method remains to be extended to particulate suspensions involving complex geometries/solid boundaries.

Lagrangian-Eulerian direct-coupling approaches have shown good potential to handle complex particle geometry and capture suspension of submicron-sized particles including Brownian effect and HI with intrinsically linear scalability. The inception of this class of approach starts by adding the so-called Landau-Lifshitz stress \citep{landau1959} to the fluid phase and coupling the subsequent fluctuating hydrodynamics (FH) method with the Newtonian dynamics for colloidal particles. The FH method explicitly captures the thermal fluctuations in the fluid phase and implicitly gives rise to the Brownian motion of suspended fine particles. Theoretical proofs \citep{Hauge1973,Mazur1982} have shown that the fluctuating stresses exerted on the particle render the particle equation of motion into a Langevin equation, in which the hydrodynamically induced random force term directly satisfies the fluctuation-dissipation theorem (FDT). Following the idea of FH, \citet{ladd1993} proposes a fluctuating lattice-Boltzmann (FLB) method to handle Brownian motion of colloidal particles. The method captures the many-body HI in both concentrated and dilute regime with O(N) scale \citep{ladd1994}. In order to satisfy the FDT, the FH-rooted method needs to directly resolve the fluid-solid interface to satisfy the no-slip boundary condition. However, this requirement poses heavy computational expense when it comes to simulating a large number of particles or concurrently handling particles with disparate length scales. To overcome this limitation, \citet{dunweg1998,dunweg1999} couples the FLB method with a MD-type approach for point particles through a simple Stokesian friction term. \textcolor{black}{Similar approaches were later developed and applied to studying the DNA translocation through nano-pores \citep{succi2007,succi2008,Succi2009}.} In this approach, thermal noises are included in both the fluid phase and particle phase. As a consequence, this method can not capture the temperature scale (thus the Brownian motion) directly and advocates an empirical rescaling of the friction coefficient. More recently, \citet{mynam2011} show that such empirical operation can be neglected by omitting the thermal fluctuation in the fluid phase while keeping the fluctuation in the particle phase. 

Following the idea of coupling fluctuating particles dynamics with non-fluctuating hydrodynamics, we develop a coupled lattice Boltzmann/Langevin dyanmics (LB-LD) approach to simulate nanoscale particle and polymer suspensions in the presence of both thermal fluctuation and many-body HI. Different from the work in \citet{mynam2011}, we couple the two phases through a forcing source distribution term \citep{he1997} that can recover the Navier-Stokes equation in the physical time scale \citep{guo2002}. Moreover, through multiple sample problems, we demonstrate that the two-way coupled LB-LD approach not only captures the Brownian motion directly (consistent with \citet{mynam2011}) but also resolves the long-range HI favorably. Through careful numerical treatment, we also demonstrate the effective applicability of the LB-LD approach to both nanoscale particle and polymer suspension problems. On the computational performance side, an Eulerian host algorithm is proposed to conduct localized neighboring particle search and interaction. This algorithm takes advantage of the existing Eulerian LB nodes and the sub-grid nature of the Lagrangian particles. With this approach, the overhead of particle dynamics (through LD) is shown to scale linearly with respect to the number of particles while adding negligible overhead to the LD framework.  Since the particle dynamics is essentially resolved in a sub-grid fashion, the LB-LD approach can be easily coupled with direct-numerical-simulation (DNS) suspension solvers to tackle multiscale, multicomponent particulate suspension flows. One example of such flows is blood flow suspended with numerous, interacting nanoscale biomolecules and microscale blood cells through microfluidic systems \citep{Ahmed2018,Griffin2018,Liu2019umiami} or biological structures \cite{Liu2018a,Liubif2018b}.

The remainder of this article is organized as follows. In \S \ref{sec:method}, the numerical method is presented. In \S \ref{sec:sample}, the accuracy and robustness of the LB-LD approach are demonstrated through multiple case studies. In \S \ref{sec:con}, we summarize and conclude the paper.

\section{Computational methods}\label{sec:method}
\subsection{Lattice-Boltzmann method}\label{sec:lb}
The method for the fluid phase with suspended particle interaction is based on the three-dimensional LB method developed in \citet{aidun1995,aidun1998,aidun2010}. The LB method solves the discretized Boltzmann equation in velocity space through the propagation of the particle distribution functions $f_i$ along the discrete lattice velocities $\mathbf{e}_i$ and the collision operation of the local distributions to be relaxed to the equilibrium distribution $f_i^{(0)}$. The collision term is simplified to the single-relaxation-time (SRT) Bhatnagar-Gross-Krook (BGK) collision operator \citep{BGK1954}, \textcolor{black}{while the more generalized multi-relaxation-time (MRT) \cite{MRT2002} and entropic \cite{yun2014a,yun2014b,KBC2014,karlin2018} collision operators can be also adopted to gain higher numerical stability.} The temporal evolution of the particle distribution function with a single relaxation time takes the form of
\begin{equation}
f_i(\mathbf{r}-\Delta t\mathbf{e}_i, t+\Delta t)=f_i(\mathbf{r},t)-\frac{\Delta t}{\tau}[f_i(\mathbf{r},t)-f_i^{(0)}(\mathbf{r},t)]+f_i^S(\mathbf{r},t),
\label{eqn:lb1}
\end{equation}
where $\tau$ is the single relaxation time scale associated with the rate of relaxation to the local equilibrium, and $f_i^S$ is a forcing source term introduced to account for the discrete external force effect \citep{he1997}. The specific formalism for $f_i^S$ is presented in \S \ref{sec:fsi}. This method has a pseudo-sound-speed of $c_s$=$\Delta r/(\sqrt{3} \Delta t)$ and a fluid kinematic viscosity of $\nu$=$(\tau -\Delta t/2)c_s^2$, where $\Delta t$ is the time step and $\Delta r$ is the unit lattice distance. The positivity of $\nu$ requires $\tau$$>$$\Delta t/2$. In the LB method, time and space in Equation (\ref{eqn:lb1}) are normalized by $\Delta t$ and $\Delta r$ such that $\Delta t_{LB}$=$\Delta r_{LB}$=1 are applied to advance the system. Details of the mapping between dimensional units and LB units are discussed in \S \ref{sec:mapping}. In the near incompressible limit (i.e., the Mach number, $Ma$=$u/c_s$$\ll$1), the LB equation recovers the Navier-Stokes equation \citep{junk2005} with the equilibrium distribution function in terms of local macroscopic variables as
\begin{equation}
f_i^{(0)}(\mathbf{r},t)=\omega_i \rho[1+\frac{1}{c_s^2}(\mathbf{e}_i\cdot \mathbf{u})+\frac{1}{2c_s^4}(\mathbf{e}_i\cdot \mathbf{u})^2-\frac{1}{2c_s^2}(\mathbf{u}\cdot \mathbf{u})],
\label{eqn:lb2}
\end{equation}
where $\omega_i$ denotes the set of lattice weights defined by the LB stencil in use. The macroscopic properties such as the fluid density, $\rho$, velocity, $\mathbf{u}$ and pressure $p$ can be obtained via moments of the equilibrium distribution functions:
\begin{subequations}
\begin{equation}
\sum_{i=1}^Q f_i^{(0)}(\mathbf{r}, t)=\rho,
\end{equation}
\begin{equation}
\sum_{i=1}^Q f_i^{(0)}(\mathbf{r}, t) \mathbf{e}_i=\rho\mathbf{u},
\end{equation}
\begin{equation}
\sum_{i=1}^Q f_i^{(0)}(\boldsymbol{r},t) \mathbf{e}_i\mathbf{e}_i=p\mathbb{I}+\rho\mathbf{u}\mathbf{u},
\end{equation}
\label{eqn:lb3}
\end{subequations}
where $\mathbb{I}$ is the identity tensor. The current study adopts the D3Q19 velocity set; that is 3 dimensions and 19 discrete velocity vectors, i.e., $Q=19$. Along the rest, non-diagonal, and diagonal lattice directions, $\omega_i$ is equal to 1/3, 1/18, and 1/36, and $|\mathbf{e}_i|$ is equal to 0, $\Delta r/\Delta t$, and $\sqrt{2}(\Delta r/\Delta t)$, correspondingly. The LB method is extensively validated \citep{aidun1995,aidun1998,DingAidun2000,aidun2010} and proved to be suitable for the direct numerical simulation (DNS) of dense suspensions of both rigid particles and deformable capsules in complex flows with good efficiency and scalability \citep{aidun1998new,ClausenJFM2011,ReasorJFM2013,ClausenCPC2010,aidun2010}.

\subsection{Langevin-dynamics method}\label{sec:ld}
\subsubsection{Governing equation}
Particles suspended in a fluid system are subjected to the impacts of the randomly fast-moving liquid molecules. When particle size is below micron-scale, such instantaneously fluctuating momentum transferred from the solvent molecules spurs the particle to yield irregular movements, known as the Brownian motion. The dynamics of such Brownian particles can be described via the Langevin equation (LE),
\begin{equation}
m_p^i\frac{d \mathbf{u}_p^i}{dt}=\mathbf{C}_p^i+\mathbf{F}_p^i+\mathbf{S}_p^i,
\label{eqn:le}
\end{equation}
where $m_p$ is the mass of the particle of index $i$. Provided the particle’s initial position, $r_{p,0}$, the displacement of the particle can be updated by integrating the particle velocity with respect to time through $\mathbf{r}_p^i$=$\mathbf{r}_{p,0}^i$+$\int \mathbf{u}_p^i dt$.

The right-hand-side (RHS) of Equation (\ref{eqn:le}) can be decomposed into three systematic forces that drive the motion of the particle. The conservative force, $\mathbf{C}^i_p$, specifies the interparticle or particle-surface interaction force that exerted on particle $i$. It is often approximated as a linear superposition of the directional derivatives of the pairwise potentials as
\begin{equation}
\mathbf{C}_p^i=-\sum_{j=0;\ j\neq i}^{N-1}\frac{d U(R_{ij})}{dR_{ij}}\frac{\mathbf{R}_{ij}}{R_{ij}},
\label{eqn:leC}
\end{equation}
where $U(R_{ij})$ is the pairwise inter-particle potential, and $\mathbf{R}_{ij}$ is a directional vector, $\mathbf{R}_{ij}=\mathbf{r}_p^i-\mathbf{r}_p^j$, connecting particles $i$ and $j$. The ingredients of $U(R_{ij})$ carry different formalism depending on the physical origins of the potential forces, which is discussed in detail in \S \ref{sec:ppinter}. The frictional force $\mathbf{F}_p^i$ is assumed to be proportional to the relative velocity of the particle with respect to the local viscous fluid [18],
\begin{equation}
\mathbf{F}_p^i=-\zeta[\mathbf{u}_p^i(t)-\mathbf{u}(\mathbf{r}_p^i,t)],
\label{eqn:leF}
\end{equation}
where $\mathbf{u}_p$ denotes the particle velocity, and $\mathbf{u}(\mathbf{r}_p,t)$ is the interpolated fluid velocity at the position where the center of the particle resides. Equation (\ref{eqn:leF}) ensures the Galilean invariance of the particle-fluid system. The details on calculating $u(\mathbf{r}_p,t)$ through interpolation are illustrated in \S \ref{sec:fsi}. The friction coefficient, $\zeta$, is determined by the Stokes’ drag law,
\begin{equation}
\zeta=3\pi \mu d_p \psi, 
\label{eqn:stokes}
\end{equation}
where $\mu$ is the dynamic viscosity of the liquid, and $\psi$ is the particle shape factor that is set to one in this study to account for spherical shape effect. The stochastic force term, $\mathbf{S}_p$, implicitly accounts for the thermal fluctuation of the solvent, and explicitly gives rise to the Brownian motion of the particle. Through the equipartition principle and the integration of the Langevin equation \citep{guazzelli2011}, the stochastic force can be related to the friction, reflecting a balance between the random thermal fluctuation and the frictional dissipation, i.e., the FDT \citep{kubo1966}. Specifically, the Cartesian component of the stochastic force exhibiting a zero mean with the covariance determined by the FDT, which reads
\begin{subequations}
\begin{equation}
\langle S_{p,\alpha}^i(t)\rangle=0,
\end{equation}
\begin{equation}
\langle S_{p,\alpha}^i(t)S_{p,\beta}^j(t)\rangle=2k_BT\zeta \delta_{ij}\delta_{\alpha\beta}\delta(t-t'),
\label{eqn:leSS}
\end{equation}
\end{subequations}
where $\alpha,\beta\in \{x,y,z\}$, $i$ and $j$ run through all the particle indices, $\delta_{ij}$ and $\delta_{\alpha\beta}$ are Kronecker deltas, $\delta(t-t')$ is the Dirac-delta function, $k_B$ is the Boltzmann constant, $T$ is the absolute temperature of the fluid bath, and the angle brackets denote the average over the ensemble of realizations of the random variables. Equations (\ref{eqn:leSS}) statistically state that the Cartesian component of $\mathbf{S}_p$ exhibits a Gaussian distribution with a zero mean. 

\subsubsection{Time scales and numerical treatment}
A colloidal system is physically enriched with multiple critical time scales, including (i) the short atomistic time scale, $\tau_a\sim10^{-12}$  sec, that is related to the frequency of rapid collisions of solvent molecules on the suspending colloidal particle, (ii) the viscous diffusion time scale,
\begin{equation}
\tau_{\nu}=\frac{d_p^2}{4\nu},
\label{eqn:letimenu}
\end{equation}
which accounts for the time for the hydrodynamic momentum to diffuse over a distance of the particle radius, (iii) the particle velocity relaxation time scale, 
\begin{equation}
\tau_{r}=\frac{m_p}{\zeta},
\label{eqn:letimer1}
\end{equation}
over which the particle velocity decays to the algebraic long-time tail regime, and (iv) the Brownian diffusion time scale,
\begin{equation}
\tau_{B}=\frac{d_p^2\zeta}{4k_BT},
\label{eqn:letimer2}
\end{equation}
which measures the time the particle has diffused its own radius. To fully resolve the HI among particles, $\tau_{\nu}$ needs to be much shorter than $\tau_B$, i.e., the Schmidt number $Sc=\tau_B / \tau_{\nu}\gg$1, allowing viscous momentum to diffuse much faster than the particle Brownian diffusion time scale \citep{guazzelli2011}. In the current study, $Sc$ lies in the range of 250$\sim$1500 for the particle size (50$\sim$300 $nm$) considered as follows. To avoid excessive computational expense when solving the LE, it is also ideal to advance the LD system with the same time step $\Delta t$ as that of the LB system. However, this requires conditional treatment of the LE to maintain stability requirement. A Stokes number, defined as $St$=$\tau_r/\Delta t$, can be introduced to characterize the relative importance of the short-time particle inertial effect. When $St<$1, i.e., the LE is advanced based on a time step greater than the particle relaxation time scale, the over-damped LE can be solved to avoid introducing sub-time steps. Under such condition, the particle motion is expected to be inertia-free and tightly follow the local fluid streamline. When $St$$\geq$1, i.e., the LE is updated with a time interval comparable to or shorter than the particle relaxation time scale, the under-damped LE needs to be solved to retain the short-time particle inertia during each time step. Under such condition, the particle tends to deviate from the streamline due to inertial effect. Since the LB evolution equation is only first-order accurate in time, a forward-differencing Euler scheme with first-order accuracy is employed to solve the discretized LE. Therefore, the velocity and displacement of the Brownian particle can be advanced, according to the St number conditions, by
\begin{subequations}
\begin{equation}
\mathbf{u}_p(t+\Delta t)=\mathbf{u}(\mathbf{r}_p,t) + \frac{1}{\zeta}[\mathbf{C}_p(t)+\mathbf{S}_p(t)], \ \ \ \ (St<1) 
\label{eqn:letledis1}
\end{equation}
\begin{equation}
\mathbf{u}_p(t+\Delta t)=\mathbf{u}_p(t)+\frac{\Delta t}{m_p}\{\mathbf{C}_p(t)+\mathbf{S}_p(t)-\zeta[\mathbf{u}_p(t)-\mathbf{u}(\mathbf{r}_p,t)]\}, \ \ \ \ (St\geq 1)
\label{eqn:letledis2}
\end{equation}
\end{subequations}
where the two discretized forms of the LE become equivalent when $St$=1. Through stability analysis, it can be shown that the condition, $|1-\Delta t\zeta/m_p|$ $\leq$1, i.e., $St$$\geq$0.5, needs to be satisfied for the discretized under-damped LE to be numerically stable. The conditional treatment as shown in equations (\ref{eqn:letledis1}-\ref{eqn:letledis2}) directly satisfies this numerical stability criterion and avoids compromising to sub-time steps \citep{dunweg1998,dunweg1999,usta2005,mynam2011}. The Gaussian distribution associated with the stochastic force $\mathbf{S}_p(t)$ is realized via a random number generator based on the Box-Muller transformation \citep{Box1958}.

\subsection{Particle-fluid coupling}\label{sec:fsi}
\begin{figure}
\centerline{
\includegraphics[width=0.8\columnwidth]{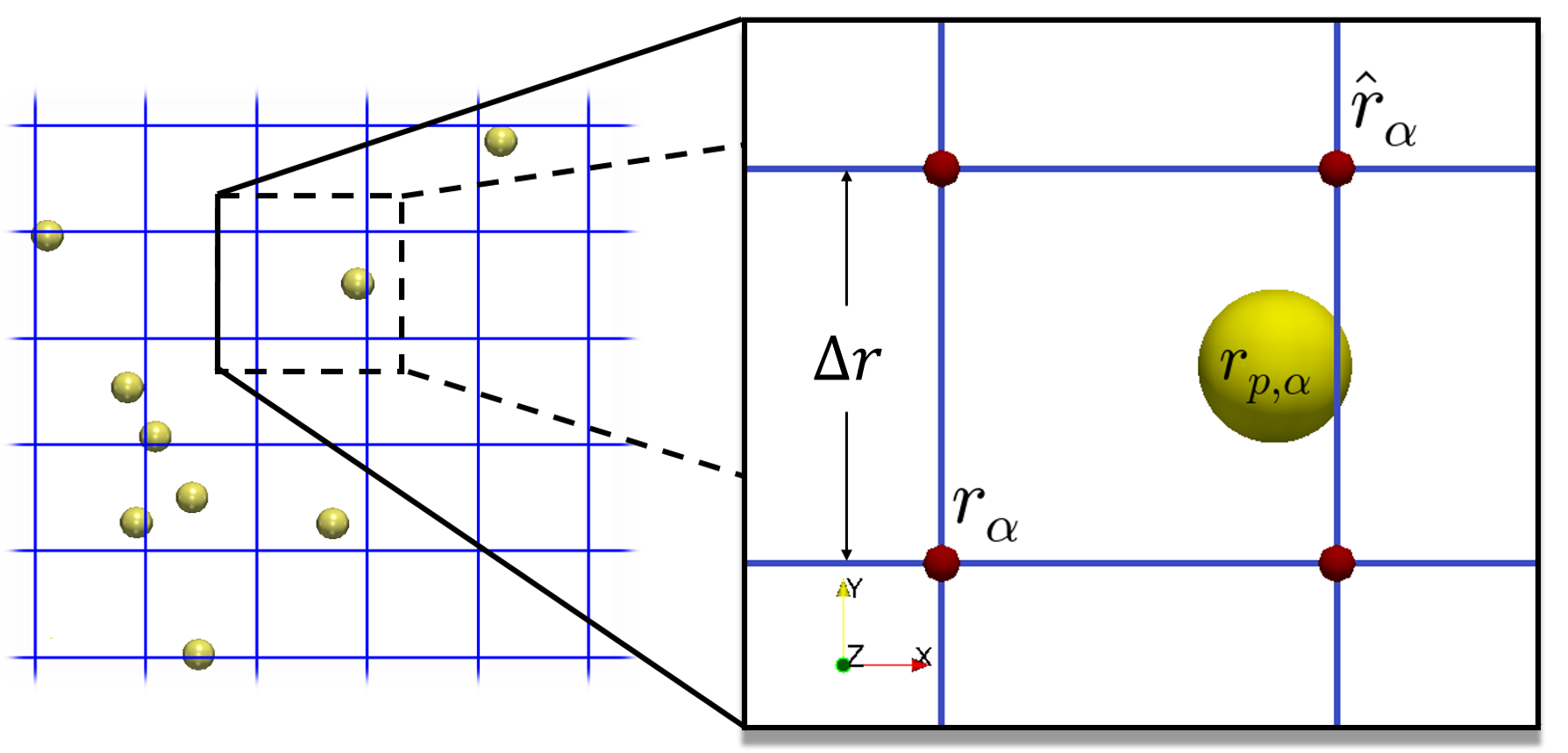}
}
\caption{2-D representation of a nanoscale particle (yellow) located at position $\mathbf{r}_p$ in a lattice cell. The lattice node $\hat{\mathbf{r}}$ is at the diagonal position of lattice node $\mathbf{r}$, and $\alpha$ denotes one of the Cartesian components.}
\label{fig:cpl}
\end{figure}

To directly capture the many-body HI mediated by the fluid phase, the interaction between the Brownian particle and the fluid is resolved by coupling the LD method to the LB method in a two-way fashion. The hydrodynamic force exerted on the particle, $\mathbf{F}_p^H$, can be systematically decomposed into a frictional component and a stochastic component \citep{dunweg1998} as
\begin{equation}
\mathbf{F}_p^H=\mathbf{F}_p+\mathbf{S}_p=-\zeta[\mathbf{u}_p(t)-\mathbf{u}(\mathbf{r}_p,t)]+\mathbf{S}_p(t),
\label{eqn:fsi1}
\end{equation}
which is applied to partially drive the particle dynamics through the LE. Meanwhile, since $\mathbf{F}_p$ and $\mathbf{S}_p$ are both originated from the collision between the particle and liquid molecules, $\mathbf{F}_p^H$ (instead of $\mathbf{F}_p$) should be assigned back to the fluid phase to conserve momentum for the entire particle-fluid system. Provided each particle or monomer is treated as a point particle and moves continuously in the lattice domain, as shown in Fig. \ref{fig:cpl}, both the construction of the fluid velocity at the center of the particle and the redistribution of the inter-phase momentum need to employ certain interpolation or extrapolation schemes.  

Two options of distance-based spatial distribution stencils with different orders of accuracy have been implemented in the current approach. For the simple trilinear stencil \citep{dunweg1998,dunweg1999}, a weight function, $w(\mathbf{r}, \mathbf{r}_p)$, can be introduced as
\begin{equation}
w(\mathbf{r},\mathbf{r}_p)=\prod_{\alpha\in\{X,Y,Z\}}\frac{|\hat{r}_{\alpha}-r_{p,\alpha}|}{\Delta r},
\label{eqn:fsi2}
\end{equation}
which is of first-order accuracy and only reads the coordinates of the first-layer, neighboring lattice nodes owned by the particle-resided lattice cell. For the stencil via Peskin’s method \citep{peskin2002}, the weight function can be introduced as
\begin{equation}
w(\mathbf{r},\mathbf{r}_p)=\prod_{\alpha\in\{X,Y,Z\}}\frac{1+cos[\frac{\pi(r_{\alpha}-r_{p,\alpha})}{4\Delta r}]}{4\Delta r},
\label{eqn:fsi3}
\end{equation}
which yields second-order accuracy and involves two layers of lattice nodes surrounding the particle. Peskin’s stencil has also been successfully applied to LB method with the external boundary force (EBF) \citep{wu2008} to resolve the curved fluid-solid boundary, which shows better accuracy and efficiency compared to the standard bounce-back (SBB) method. The following studies employ the trilinear stencil for simplicity. 
By employing the weight function, the background fluid velocity at the particle center can be interpolated as
\begin{equation}
\mathbf{u}(\mathbf{r}_p,t)=\sum_{\mathbf{r}\in N_c}w(\mathbf{r},\mathbf{r}_p)\mathbf{u}(\mathbf{r},t),
\label{eqn:fsi4}
\end{equation}
where $N_c$ denotes the group of nodes on the lattice cell occupied by the particle. The same weight function can be applied to the construction and distribution of the reactionary impulse due to the particle-fluid interactions. Specifically, the reactionary impulse density \citep{aidun1998}, during each time step, can be assigned to the surrounding lattice nodes as
\begin{equation}
\mathbf{J}(\mathbf{r},\mathbf{r}_p)=-w(\mathbf{r},\mathbf{r}_p)\frac{\mathbf{F}_p^H\Delta t}{\Delta r^3},
\label{eqn:fsi5}
\end{equation}
where $\mathbf{J}(\mathbf{r},\mathbf{r}_p)$ is the impulse density to be assigned to the lattice node, $\mathbf{r}$, due to particle-fluid interaction at the particle position, $\mathbf{r}_p$, at each time step. A local forcing source distribution term, $f_i^S (\mathbf{r},t)$, based on the method proposed in \citet{he1997} can then be calculated as
\begin{equation}
f_i^S(\mathbf{r},t)=\frac{\omega_i \mathbf{J}(\mathbf{r},\mathbf{r}_p)\cdot \mathbf{e}_i}{c_s^2}.
\label{eqn:fsi6}
\end{equation}
Instead of modifying the local equilibrium distribution functions as shown in previous studies \citep{dunweg1998,mynam2011}, the current approach, similar to the EBF method \citep{wu2008}, modifies the general LB evolution equation into Equation (\ref{eqn:lb1}) by adding the forcing distribution function $f_i^S (\mathbf{r},t)$, which is shown to approximate the Navier-Stokes equation in the physical time scale \citep{guo2002}. \textcolor{black}{The same forcing term has been applied to the studies of DNA translocation through nano-pores \citep{succi2007,succi2008,Succi2009}.} However, those studies use zeroth order spatial distribution schemes and introduce thermal fluctuation in both fluid and solid phases, which is not aligned with the LB-LD approach proposed in the current study. 

\subsection{Particle-particle interactions}\label{sec:ppinter}
\subsubsection{Interparticle potential for suspended particles}
The inter-particle interactions between unconnected particles are assumed to follow the classic Derjaguin-Landau-Verwey-Overbeek (DLVO) theory that includes both the standard van der Waals potential \cite{Everaers2003} and the electrostatic contribution to the overall DLVO potential, as advocated in \citet{Schunk2012}. For the attractive component of the van der Waals potential, the formulation of \citet{hamaker1937} is employed as follows,
\begin{equation}
U_{A}(R_{ij})=\frac{A_{cc}}{6}[\frac{d_p^2}{2(R_{ij}^2-d_p^2)}+\frac{d_p^2}{2R_{ij}^2}+\text{ln}(\frac{R_{ij}^2-d_p^2}{R_{ij}^2})],
\label{eqn:att}
\end{equation}
where $A_{cc}$ is the Hamaker constant that is set to $A_{cc}=4\pi^2k_BT$ according to \citet{Schunk2012}. The repulsive component of the van der Waals potential adopts the integrated Lennard-Jones (LJ) potential derived in \citet{Everaers2003} as
\begin{equation}
\begin{aligned}
U_{R}(R_{ij})=\frac{A_{cc}\sigma^6}{37800R_{ij}}[\frac{R_{ij}^2-7d_pR_{ij}+13.5d_p^2}{(R_{ij}-d_p)^7}+\frac{R_{ij}^2+7d_pR_{ij}+13.5d_p^2}{(R_{ij}+d_p)^7}\\-2\frac{R_{ij}^2-7.5d_p^2}{R_{ij}^7}],
\end{aligned}
\label{eqn:rep}
\end{equation}
where $\sigma$ is the repulsive scaling factor that can be set to $\sigma=d_p/10$ as suggested in previous studies \cite{Schunk2012,Bolintineanu2014}.
Together with the screened electrostatic potential, $U_E$ \cite{Schunk2012}, the total DLVO potential can be calculated as $U_{DLVO}=U_A+U_R+U_E$. In the following study (\S \ref{sec:hinder}), neutral electrostatic effects are considered, i.e., $U_E=0$. However, the charge effect can be incorporated through various electrostatic models \cite{Schunk2012,Griffin2018}.

\subsubsection{Interparticle potential for chain of particles}
In the case of polymer chains, particles (beads) are connected by elastic springs to form bead-spring chains. To account for the inter-bead cohesive effect and the bead volume-exclusion effect, the standard Lennard-Jones (LJ) potential is employed according to
\begin{equation}
U_{LJ}(R_{ij})=\tilde{\epsilon} k_BT[(\frac{d_p}{R_{ij}})^{12}-2(\frac{d_p}{R_{ij}})^6],
\label{eqn:lj}
\end{equation}
where $\tilde{\epsilon}$ is the scaling factor of the LJ potential well depth, which can be tuned to adjust the cohesiveness of the polymer chain. In the following simulations, $\tilde{\epsilon}$ is set to $1.8$ to obtain the best fit to the experimental data discussed in \S \ref{sec:vWF}. Equation \ref{eqn:lj} is truncated at a cut-off distance $R_{ij}^c$=8$d_p$ to limit the bound for neighboring bead search but still preserve the major cohesive effect between adjacent beads. The cohesive strength among beads plays a critical role in regulating the conformation of self-associable polymers, such as von Willbrand factors (vWF) \citep{Katz2006}. The inter-bead connectivity is established through a finitely extensible nonlinear elastic (FENE) spring \citep{grest1990} as
\begin{equation}
U_{S}(R_{ii'})=-\frac{2\tilde{k}k_BT\Delta R_{max}^2}{d_p^2}\text{ln}[1-(\frac{R_{ii'}-d_p}{\Delta R_{max}})^{2}],
\label{eqn:fene}
\end{equation}
where $R_{ii'}$ is the center-to-center distance between bead $i$ and its neighbor $i'=i+1$ (or $i-1$), $\tilde{k}$ is the scaling factor of the spring tensile elasticity, and $\Delta R_{max}$ denotes the maximum bond extension. Here, $\Delta R_{max}$ is chosen to be 0.25$d_p$ to limit the extension of the polymer chain \citep{dunweg1999}. The spring scaling factor, $\tilde{k}$, is set to $200.0$ as suggested in \citet{Katz2006}.  Unlike the linear Hookean connectivity model used in other studies \citep{Katz2006,schneider2007shear}, the FENE spring captures the hyperelastic-like behavior when the polymer bond elongation reaches its maximum. 

\subsection{Mapping between physical units and LB units}\label{sec:mapping}
All equations and variables thus far are introduced in physical units for consistency. To follow the convention of the LB method, the time, length, and density/mass units of the entire LB-LD system need to be mapped to LB units in the simulations \citep{aidun2010}. The mapping is performed such that $t$, $r$, and $\rho$ are normalized by $\Delta t$, $\Delta r$, and $\rho$, respectively. Therefore, the unit lattice distance, time step, and fluid density in LB units are obtained as $\Delta r_{LB}$=$\Delta t_{LB}$=$\rho_{LB}$=1. The normalized single relaxation time scale and the corresponding LB viscosity are  $\tau_{LB}$=$\tau/\Delta t$ and $\nu_{LB}$=$(2\tau_{LB}-1)/6$, respectively. As mentioned in \S \ref{sec:lb}, the LB single relaxation time, $\tau_{LB}$, needs to satisfy $\tau_{LB}>$0.5 to produce positive viscosity. 

The current study selects mapping ratios of $\frac{\Delta t}{\Delta t_{LB}}$=$\frac{\nu_{LB}}{\nu}(\frac{\Delta r}{\Delta r_{LB}})^2$ for time, $\frac{\Delta r}{\Delta r_{LB}}$=333 $nm$ for length, and $\frac{\rho}{\rho_{LB}}$=1000 $kg/m^3$ for density. The time mapping ratio depends on the fluid kinematic viscosity and the LB single relaxation time in use. The fluid density is selected to be 1000 $kg/m^3$ and the viscosity 1.2 $cP$. The temperature is set to $T=310\ K$. It should be noted that the current mapping strategy is based on the SRT LB method. \textcolor{black}{However, more flexibility can be obtained to match a broader spectrum of fluid and thermal properties with the MRT and entropic LB method \cite{MRT2002,yun2014a,yun2014b,KBC2014}.} For clarity, the LB counterparts of previously introduced quantities in physical units are denoted with the subscript `LB' hereafter.

\subsection{Eulerian-host algorithm}\label{sec:EHA}
\begin{figure}
	\centerline{\includegraphics[width=0.9\columnwidth]{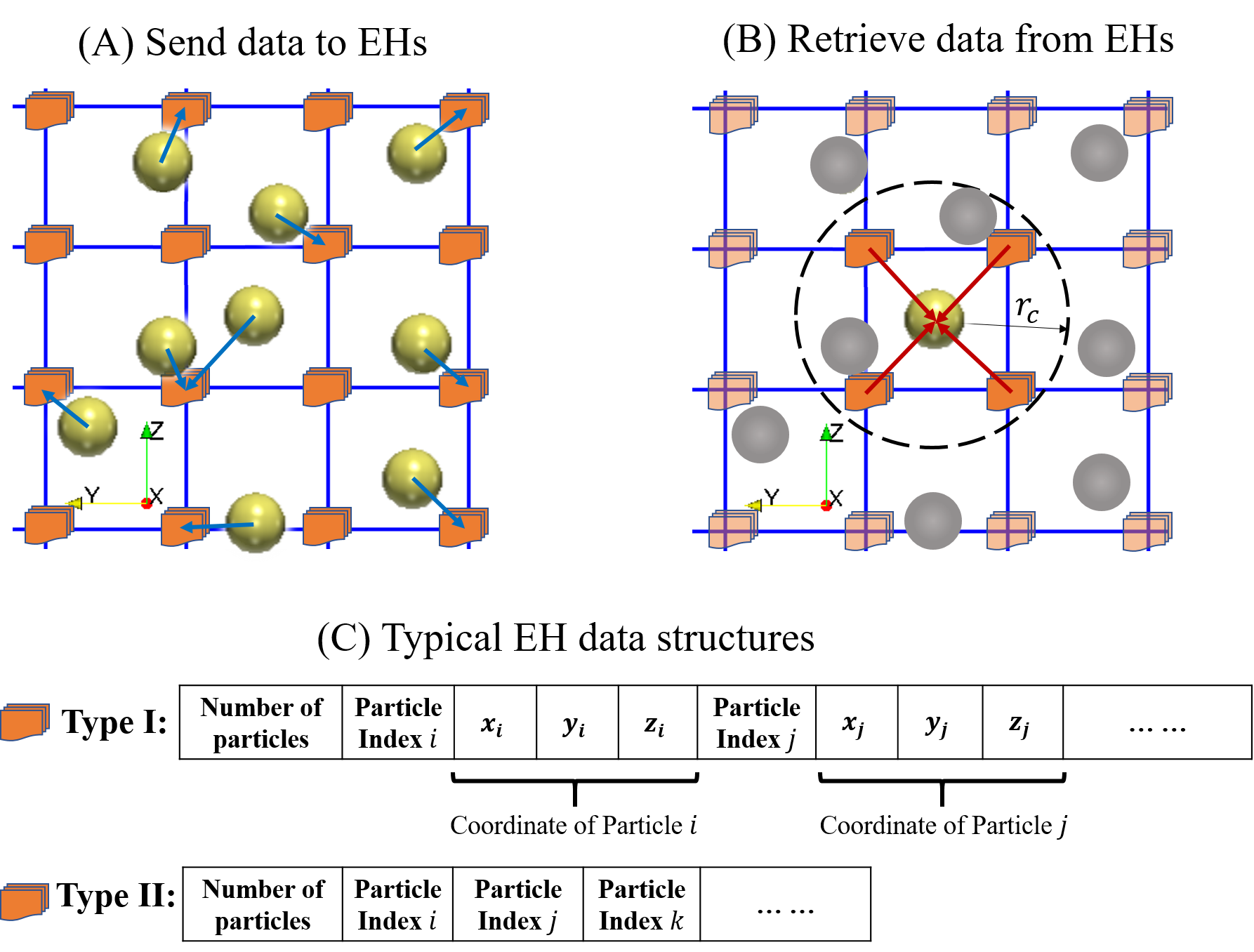}}
	\caption{Schematics of the EH algorithm: (A) Once an updated particle position is obtained, the coordinates of the particle is sent to its nearest EH; (B) A single particle interacts with its neighboring EHs that contain the coordinates of the adjacent particles; (C) Typical date structures of the Eulerian host.}
	\label{fig:EH1}
\end{figure}
Previous hybrid approaches for simulating particle suspensions have been reported to scale linearly with respect to the number of particles, $N$, \citep{dunweg1998,dunweg1999,mynam2011}. However, when simple search (SS) algorithm is applied to unconnected particle-particle interactions, the scaling of the particle dynamics could still degrade to $O(N^2)$, which consequently undermines the overall scalability of the method. To overcome this issue, two types of algorithms have been proposed in the MD community. The Verlet neighbor list (VNL) algorithm \citep{verlet1967} limits the pair search to a list of $N_{nl}$ adjacent particles for a given particle, which reduces the search scaling to $O(NN_{nl})$; however, the construction of the neighbor list still scales as $O(N^2)$, which undermines the overall efficiency when $N$ becomes reasonably large. The cell linked list (CLL) algorithm \citep{hockney1988} partitions the simulation domain into cellular domains, and each particle only interacts with the particles within the same cell. The CLL algorithm truly scales as $O(N)$ but requires extra computational infrastructure to handle domain decomposition.

Inspired by the merit of both VNL and CLL algorithm and noticing the Eularian nature of the LB method, we propose a linear short-range particle-pair search algorithm that makes use of the Eulerian LB nodes as information hosts for the Lagrangian particles. As illustrated in Figure \ref{fig:EH1} (A), during each LB time step, the updated information of each sub-grid particle can be stored in a data structure on its nearest LB node, which is denoted as an Eulerian host (EH). When performing particle-particle short-distance interactions, each particle only interacts with its surrounding EHs that carry the information of the neighboring particles within a cut-off distance, as illustrated in Figure \ref{fig:EH1} (B). Each EH could contain information such as the number of particles, the index of each particle and even the coordinate of each particle, as indicated in Figure \ref{fig:EH1} (C). The complexity of the data structure for EHs varies depending on the concentration of the sub-grid particles. Specifically, when the particle volume fraction is low, i.e., each EH only contains 1$\sim$2 particles, data structure of type I can be used with the information of particle coordinates included; when the particle volume fraction is high, i.e., each EH could contain more than two particles, data structure of type II can be used to avoid excessive memory allocation for each LB node. The benefit of using EHs of type I is that the particle information can be directly communicated together with the LB fluid node information during MPI permutations. Therefore, no separate MPI communications are needed for the particle phase. However, when type II EH is applied, separate MPI communications for the particle phase are necessary since the particle coordinates are stored separately from the LB fluid nodes information. For the case with dilute particle concentrations, EHs of type I data structure are adopted.

\begin{table}[bt]
\centering
\begin{tabular}{ccccc}
\hline
\textbf{Number of Particles, N} & \textbf{0} & \textbf{100} & \textbf{1\ 000} & \textbf{10\ 000}\\
TWCT (s), EH & 464.3 & 470.3 & 522.2 & 1\ 050.7\ \ \\
TWCT (s), SS & 300.7 & 307.2 & 565.8 & 21\ 857.7\ \ \\
LD overhead (s), EH & 0 & 6.0 & 57.9 & 568.5\ \ \\
LD overhead (s), SS & 0 & 6.4 & 265.1 & 8\ 442.4\ \ \\
\hline  
\end{tabular}
\caption{Scaling performance comparison between the Eulerian-host algorithm and the simple search algorithm.}
\label{tab:1}
\end{table}
\begin{figure}
\centerline{\includegraphics[width=0.8\columnwidth]{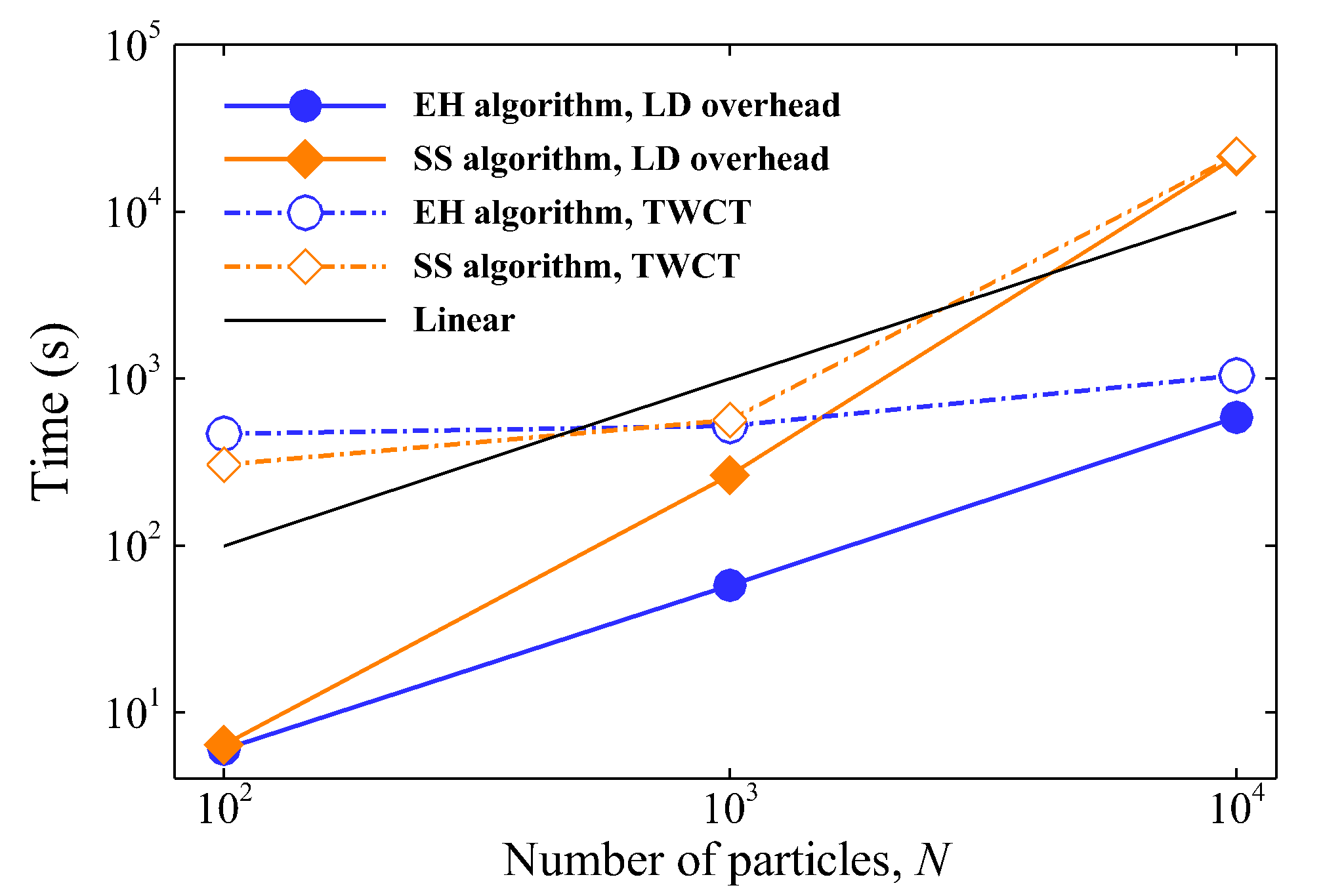}}
\caption{The Langevin-dynamics (LD) overhead and the total wall clock time (TWCT) using the Eulerian-host (EH) algorithm or simple-search (SS) algorithm plotted against the number of particles in log-log scale. The EH algorithm is shown to scale linearly with respect to the number of particles in contrast to the quadratic scale of the SS algorithm.}
\label{fig:EH2}
\end{figure}
\begin{figure}
\centerline{\includegraphics[width=0.9\columnwidth]{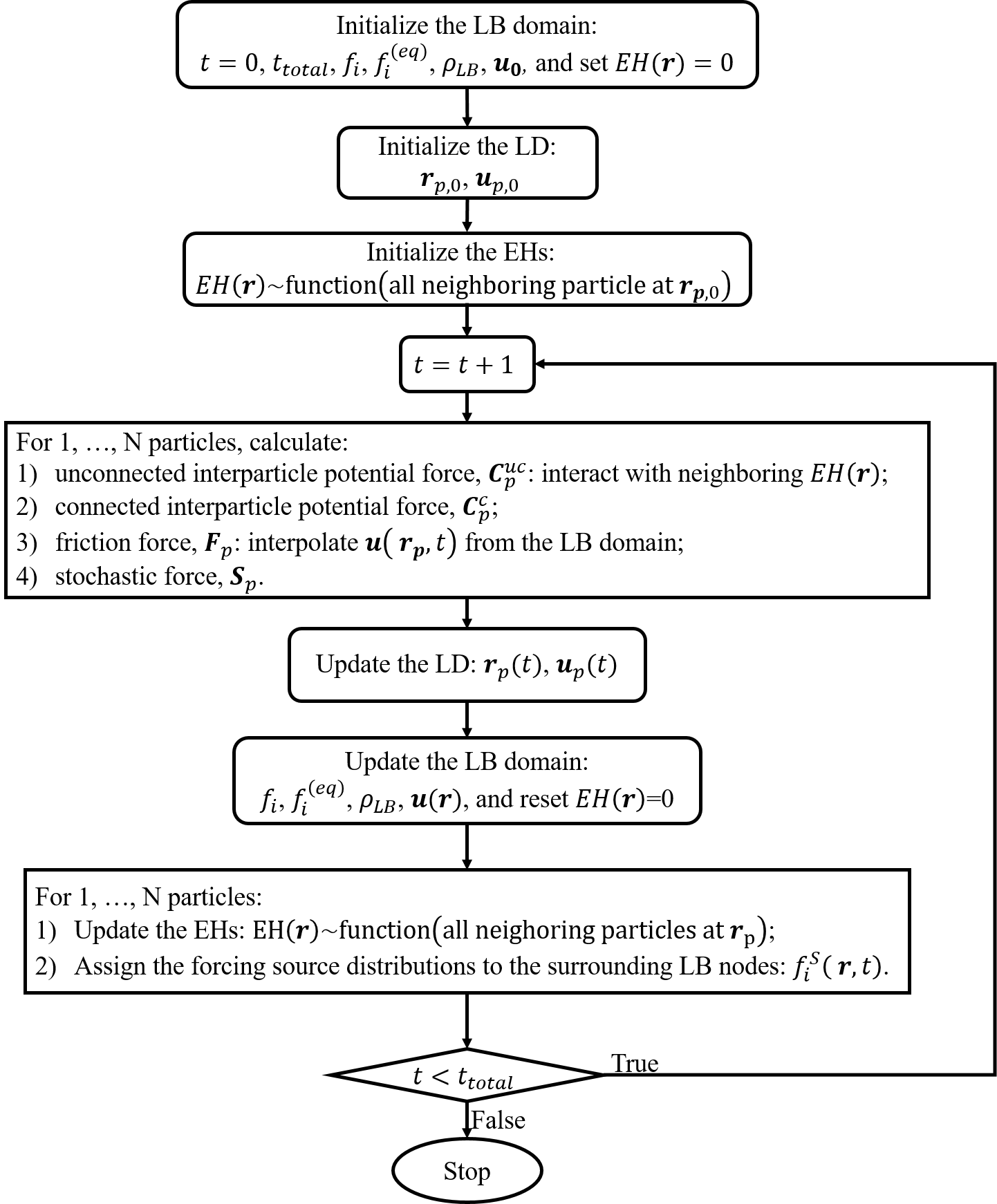}}
\caption{The flow chart of the current LB-LD method embedded with the EH algorithm to handle short-distance, unconnected particle-particle interaction.}
\label{fig:flowchart}
\end{figure}

The particle-number scalability of the LB-LD approach embedded with the EH algorithm is tested by simulating different number ($N$=0, 100, 1000, and 10000) of particles under a wall-bounded shear flow within a $50^3$ LB domain. For each case, five runs are sampled to obtain an averaged total wall clock time. All the cases are tested for 10\ 000 LB steps in serial mode on the TACC Stampede-2 supercomputer. Each computing node is configured by one Intel Xeon Phi 7250 processor, which has a base frequency 1.4 GHz and contains 68 cores. The total wall clock time (TWCT) for each case and the corresponding overhead associated with LD operations are tabulated in Table \ref{tab:1}. The LD overhead for cases with $N>0$ is calculated by subtracting the corresponding TWCT with that of the $N$=0 case. As plotted in Figure \ref{fig:EH2}, the EH algorithm compared to the SS algorithm shows slightly higher TWCT when $N$$\leq$100. This is understandable since the introduction of EHs adds overhead to the update of the Eulerian LB domain. However, as the number of particles increase, the EH cases starts to show much lower TWCT compared to the SS cases. In fact, when $N$=10\ 000, the TWCT with EH algorithm becomes more than one magnitude lower than that with the SS algorithm. A linear scalability curve, generated by setting the overhead equal to the number of particles, is also plotted as a reference. The results clearly show that the LD overhead with EH algorithm scales almost linearly with the number of particles $\sim$$O(N)$, while the simple search algorithm exhibits $\sim$$O(N^2)$ scales. 

Both the construction of EHs and the interaction with EHs are localized, therefore scaling as $O(N)$. For interactions among 100 $nm$ NPs, a search of neighboring 8 EHs is shown to be enough to cover all the nearby particles within the cut-off distance. As the particle size increases, a slight expansion of the search range is needed, which however doesn’t undermine the $O(N)$ merit. When the particle is encountered with a periodic boundary, a wrapping operation is needed to ensure the availability of EHs. By applying the EH algorithm, the particle-particle interaction can be conveniently handled together with the particle-fluid coupling procedures. The EH algorithm provides the localized data structure for the particle dynamics of Lagrangian nature, which is fundamentally more convenient for parallelization. Figure \ref{fig:flowchart} depicts the flow chart of the current LB-LD method with the EH algorithm embedded with the LB-LD two-way coupling scheme. Both the rectangular blocks in Figure \ref{fig:flowchart} denote the processes of force calculation and particle-fluid/particle interaction associated with each particle, which are purely localized operations and scale as $O(N)$.

\section{\textcolor{black}{Model verification and validation}}\label{sec:sample}
The validity and accuracy of the LB-LD approach to capturing the dynamics of nanoscale particle and polymer suspensions subject to both the thermal fluctuation and HI are demonstrated through several sample problems. First, the momentum relaxation of an isolated particle is presented to show the correctness of the particle-fluid coupling. Then, the self-diffusion of colloidal particles in infinite dilution is demonstrated to show the direct capture of Brownian motion. As follows, the hindrance of particle diffusion in concentrated colloidal suspension is discussed to shown the validity of the short-distance particle-particle interaction model. The self-globularization process of a cohesive polymer chain and the shear-induced unfolding of a collapsed polymer globule are further presented to show the applicability of the LB-LD model to nano-polymer suspension dynamics subject to HI effects. \textcolor{black}{All cases adopt a LB relaxation time $\tau_{LB}$=1 unless otherwise prescribed.}

\subsection{Momentum relaxation of an isolated particle}
The fluid-particle coupling is first verified by analyzing the slowing-down process of an isolated particle with an initial momentum in a quiescent viscous fluid. \textcolor{black}{A cubic LB domain with periodic boundary condition enforced in each direction is selected for all the simulations. Three computational domains with different dimensions ($50^3$, $70^3$, and $100^3$) are considered to study the domain size dependency.} A particle of mass $m_{p,LB}$=29.3 with an initial velocity $u_{p,LB}$ (0)=0.01 along the $\text{X}$ direction is released in the center of the domain. \textcolor{black}{By setting $\tau_{LB}$=0.51 and $d_p$=100 $nm$}, a friction coefficient $\zeta_{LB}$=0.48 is prescribed to dissipate the kinetic energy of the particle. Given $St>1$, under-damped LE is solved for this problem. 

\begin{figure}
\centerline{
\includegraphics[width=0.5\columnwidth]{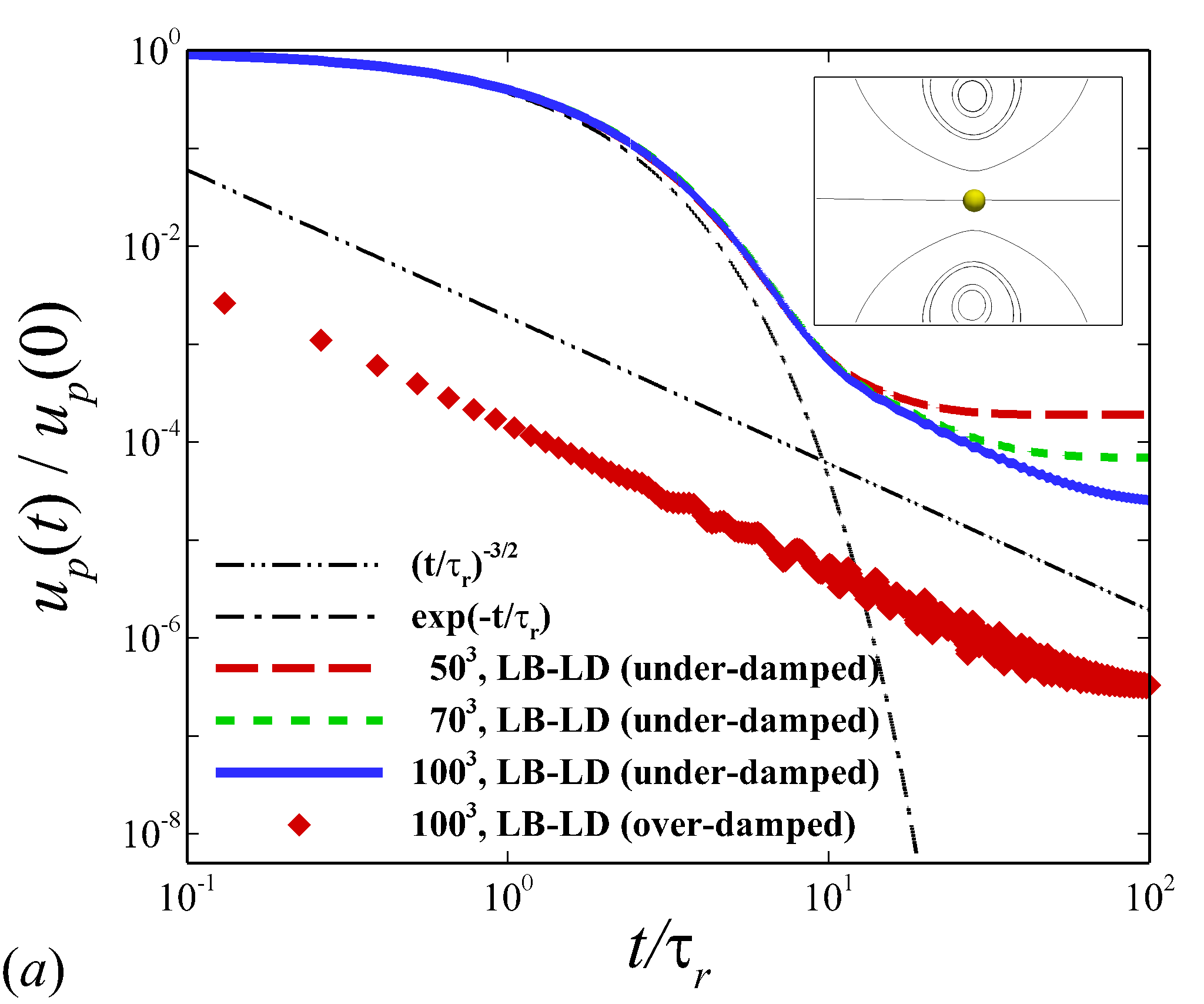}
\includegraphics[width=0.5\columnwidth]{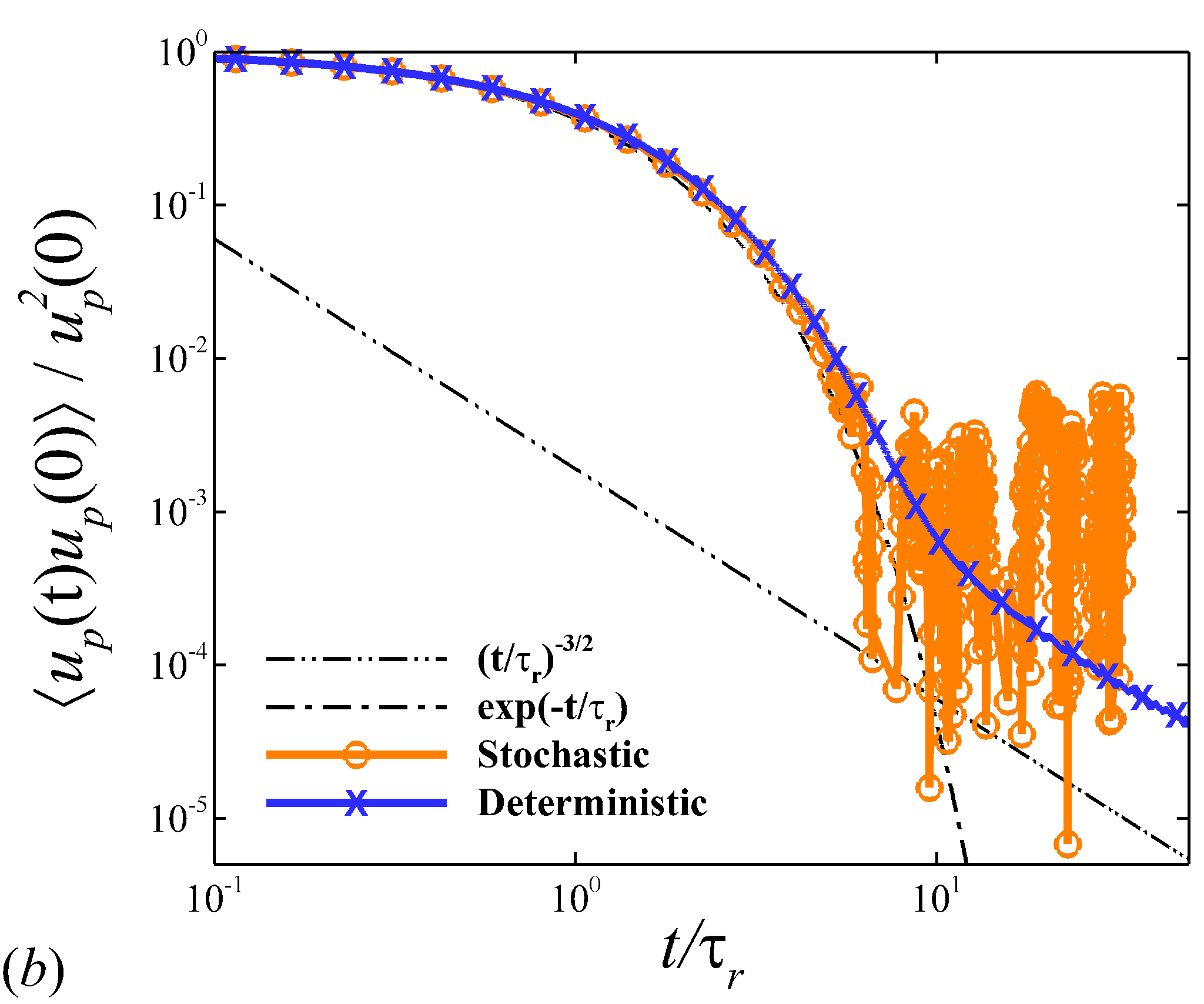}
}
\caption{The momentum relaxation process of a point particle with an initial disturbance in periodic LB fluid domains of different sizes (a) under deterministic condition or (b) subject to stochastic noise effects.}
\label{fig:vaf}
\end{figure}

The deterministic response of the particle momentum relaxation is first presented by omitting the Brownian effect. The inset of Figure \ref{fig:vaf} (a) shows a snapshot of the flow field induced by the decelerating particle right after its release. Such double vortex flow structure has also been observed in previous numerical studies \citep{alder1970,ladd2001}. The time evolution of the particle velocity normalized by the initial velocity for three domain sizes are depicted in Figure \ref{fig:vaf} (a), where the corresponding asymptotic behaviors are also presented for comparison. At short-time scales, i.e., when $t/\tau_r<10$, the particle velocity decays exponentially for all three domain sizes, which also agrees well with the asymptotic short-time exponential decay behavior, $exp(-t/\tau_r)$. At long-time scales, \citet{alder1970} show the particle velocity should eventually decay according to a power law scale, $(t/\tau_r)^{-3/2}$, known as the long-time tail behavior, which reflects the fluid-particle coupling effect. As shown in Figure \ref{fig:vaf} (a), the long-time tail behavior is not significant for the $50^3$ domain size, where instead a plateau is observed right subsequent to the exponential decay. However, as the domain size increases to $100^3$, the long-time tail behavior appears to be more pronounced. The eventual flattening of all cases manifests the fluid and particle eventually translate at the same velocity as a result of the periodicity of the finite LB domain and the conservation of momentum for the particle-fluid system.
The same problem with domain size of $100^3$ is further simulated using the over-damped LE with zero particle inertia, as also depicted in Figure \ref{fig:vaf} (a). As expected, the particle velocity directly relaxes to the long-time tail regime without yielding the exponential decay behavior. The jittering of the relaxation curves for the zero-inertia case is due to the temporal discretization effect and can be eliminated by reducing the time step \citep{dunweg1998,mynam2011}.

According to the FDT, the stochastic relaxation behavior of a Brownian particle should be in accordance with the deterministic relaxation behavior of the particle under the same flow condition with no Brownian effect. This can be demonstrated by adding the Brownian noise to the previous deterministic case. For simplicity, the stochastic force is only applied to the particle along the +X direction, thus the motion of particle is constraint in a one-dimensional fashion. The particle velocity relaxation process can be quantified by the normalized velocity autocorrelation function (VAF), $\langle u_p(t)u_p(0)\rangle/u_p^2 (0)$, which reduces to $u_p(t)/u_p(0)$ for the deterministic case. In Figure \ref{fig:vaf} (b), the particle VAF for cases with or without Brownian effect is plotted against the LB time. The VAF curve for the stochastic case is obtained by averaging over an ensemble of ten independent runs. Overall good agreements between the stochastic case and the deterministic case are obtained. These results show the particle-fluid coupling and the FDT are correctly captured using the current LB-LD method.

\subsection{Brownian diffusion in dilute colloidal suspensions}
\begin{figure}
\centerline{
\includegraphics[width=0.5\columnwidth]{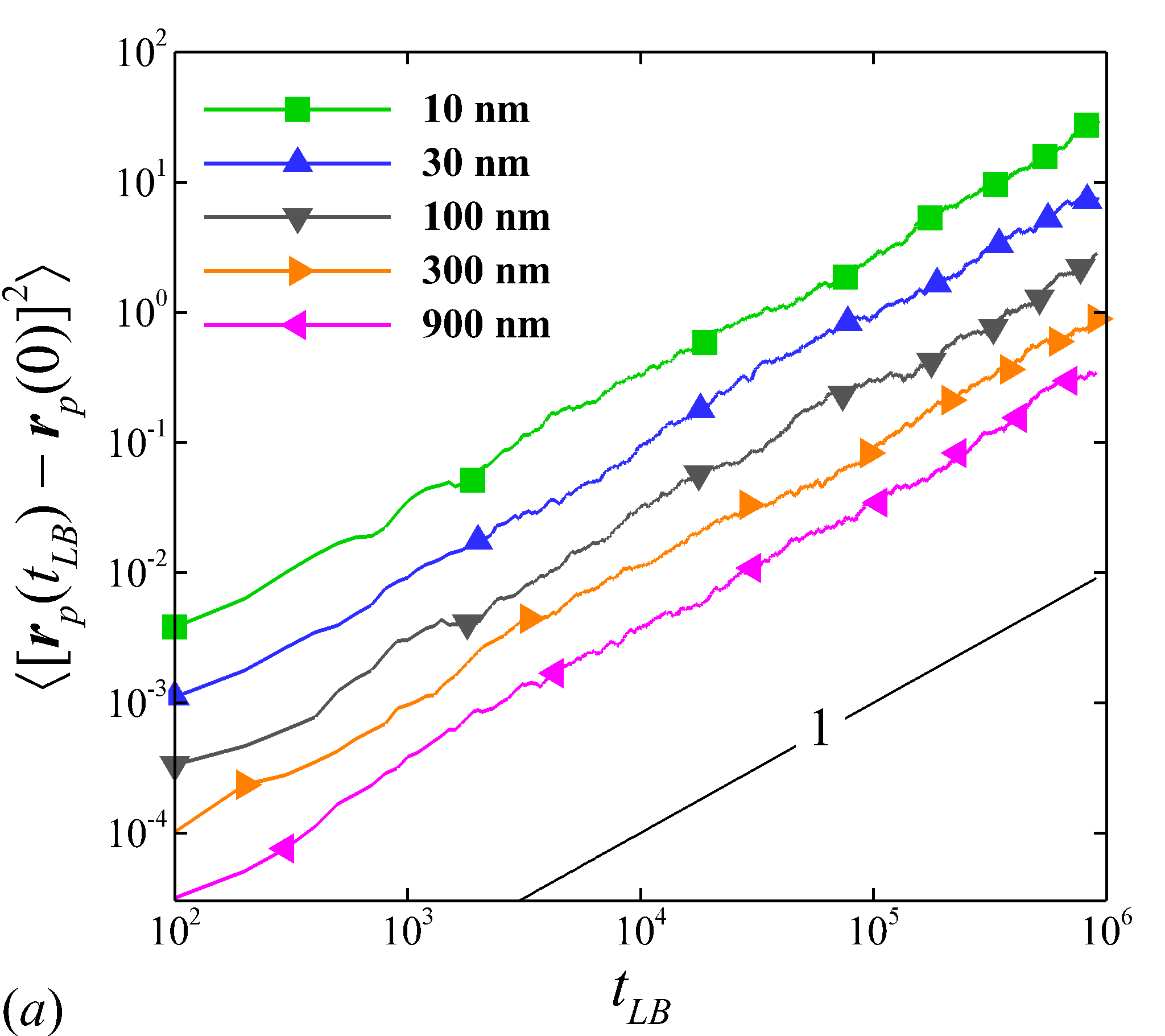}
\includegraphics[width=0.5\columnwidth]{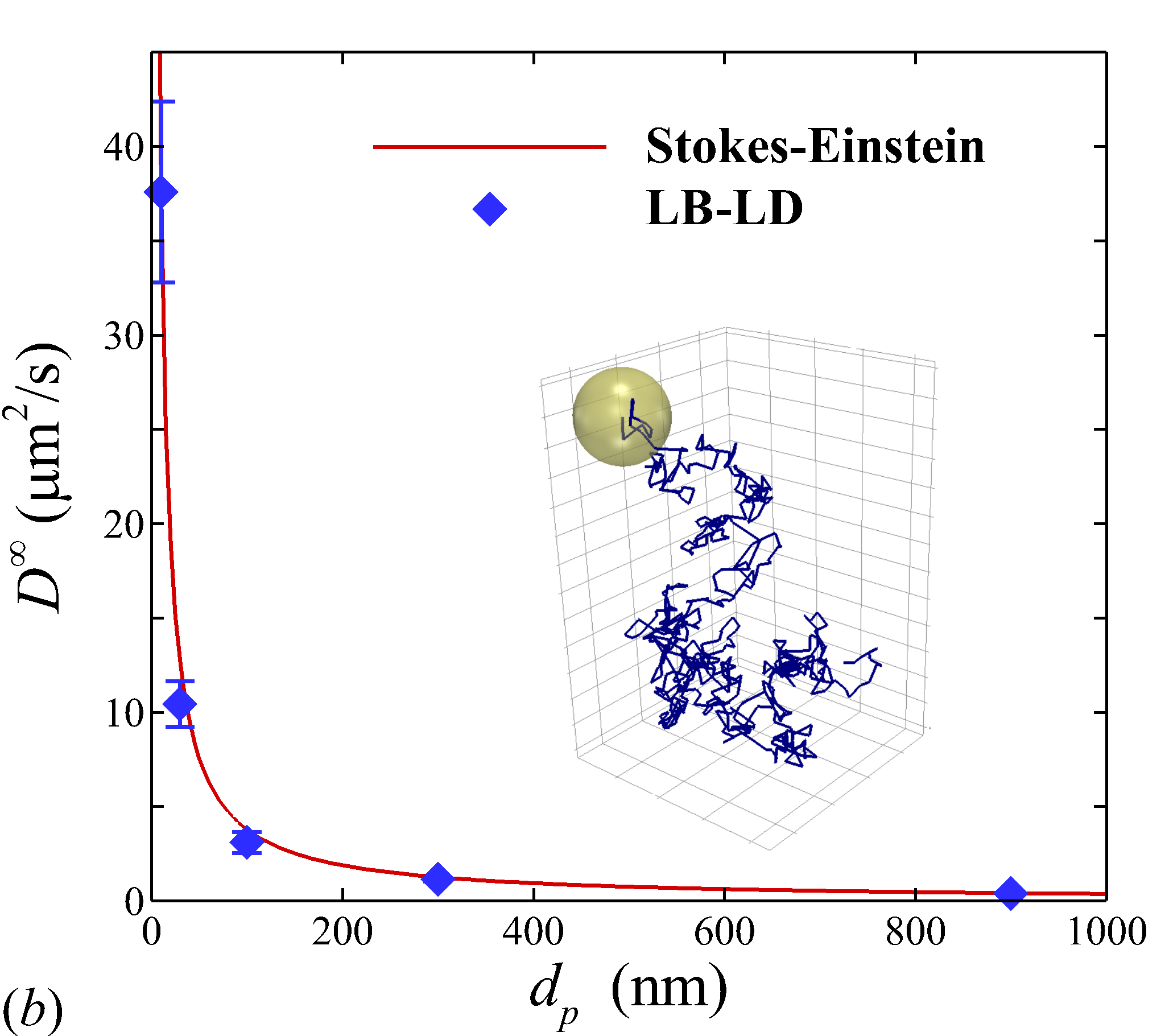}
}
\caption{(a) The mean squared displacements for various particle sizes in dilute concentraions. (b) Normalized self-diffusivity of colloidal particles as a function of particle size at infinite dilution. The symbols are the simulation results. The line is the theoretical results based on the Stokes-Einstein relation.}
\label{fig:diff_d}
\end{figure}
Previous analysis \citep{Hauge1973,Mazur1982} show that once the fluid phase satisfies the FDT through FH, the particle dynamics is automatically rendered to yield the FDT through the particle-fluid coupling. However, when both the particle and fluid phase are introduced with fluctuation that satisfies the FDT individually \citep{dunweg1998,dunweg1999}, the coupled system requires an empirical rescaling of the particle mobility to produce the correct temperature and hence the Brownian effect. Below, we study the self-diffusion of colloidal particles in the dilute regime to show that the current LB-LD approach, which introduces thermal fluctuation solely to the particle phase, directly captures the correct Brownian effect. 

Colloidal suspensions in the dilute regime with five particle sizes, $d_p=10, 30, 100, 300$ and $900\ nm$, are simulated independently to evaluate the long-time particle diffusivity. Since only the long-time diffusion is of concern, the over-damped LE is employed in this study. By sampling the mean-squared displacement (MSD) of the Brownian particle, the particle long-time diffusivity, $D^{\infty}$, can be calculated as
\begin{equation}
D^{\infty}=\frac{1}{6}\frac{d}{dt}\langle[\mathbf{r}_p(t)-\mathbf{r}_p(0)]^2\rangle |_{t\rightarrow \infty},
\label{eqn:msd1}
\end{equation}
at $t/\tau_r\gg 1$, where the angle brackets denote an ensemble of 1\ 000 particles in one simulation. \textcolor{black}{All cases are simulated in a periodic cubic domain with dimensions of $100^3$. This setup ensures the particle volume concentration being kept below 0.1$\%$.} The theoretical Brownian diffusion in dilute, unconfined suspension systems is determined by the Stokes-Einstein relation as,
\begin{equation}
D^B=\frac{k_BT}{\zeta},
\label{eqn:msd2}
\end{equation} 
which is a nice, compact manifestation of the FDT by quantitatively relating the particle thermally induced diffusive behavior to the dissipative property of the particle-fluid system.

Figure \ref{fig:diff_d} (a) shows the transient growth of the particle MSD for different particle size. All the cases exhibit linear temporal growth of the MSD due to the neglecting the of the particle inertia. Besides, as the particle size decreases, an increase of the MSD is observed at the same instance. Figure \ref{fig:diff_d} (b) further plots the dimensional long-time particle diffusivity versus the particle size. The inset in Figure \ref{fig:diff_d} (b) shows a typical example of the particle trajectory. The particle long-time diffusivity predicted by the LB-LD simulation shows excellent agreement with the Stokes-Einstein theory. This observation proves that the LB-LD approach, which only introduces thermal fluctuation in the particle phase, directly captures the Brownian motion without the any empirical adjustment.

\subsection{Hindered particle diffusion in concentrated colloidal suspensions}\label{sec:hinder}
In concentrated colloidal particle suspensions under quiescent flow condition, the particle self-diffusion is often hindered due to the frequent, short-distance particle-particle interactions. The hindered particle long-time diffusivity, $D^{\infty}$ is known to be sensitive to the number of particles, $N$, considered in the system \citep{ladd1990}. To correct for such finite-size effects, \citet{ladd1990} proposes a correction of the long-time diffusivity to link the $N$-particle system to an infinite-particle system through
\begin{equation}
\frac{D^{\infty}}{D^B}=\frac{D^{t}(N)}{D^B}+\frac{\mu}{\mu_{\phi}}[1.7601(\frac{\phi}{N})^{1/3}-\frac{\phi}{N}],
\label{eqn:hinder}
\end{equation}
where $D^{t}(N)$ is the particle self-diffusivity measured with a $N$-particle system at time $t$, $\mu$ is the viscosity of the pure liquid and $\mu_{\phi}$ is the viscosity of the particle suspension at various concentrations. To further validate the LB-LD model, we simulated concentrated colloidal suspension with particle size of $d_p=600\ nm$ and particle packing fractions ranging from $\phi=$0 to 42$\%$. Three problem sizes are considered with particle number $N=1073, 1637$ and $2096$. \textcolor{black}{The periodic compuational cube for each setup is adjusted according to the particle packing fraction.} The long-time diffusivity is measured at $t/\tau_B=1.0$, which is shown to be the least time needed in order to reach the long-time diffusivity plateau \cite{Schunk2012}. It should be noted that the $t/\tau_B=1.0$ criteria, somewhat empirical, might need future verification to ensure the full long-time regime arrived. The measured diffusion coefficients, $D^{t}(N)$, are further corrected based on Equation \ref{eqn:hinder}, where the suspension viscosity adopts the values reported in \citet{Bolintineanu2014}. 
\begin{figure}
\centerline{
\includegraphics[width=0.5\columnwidth]{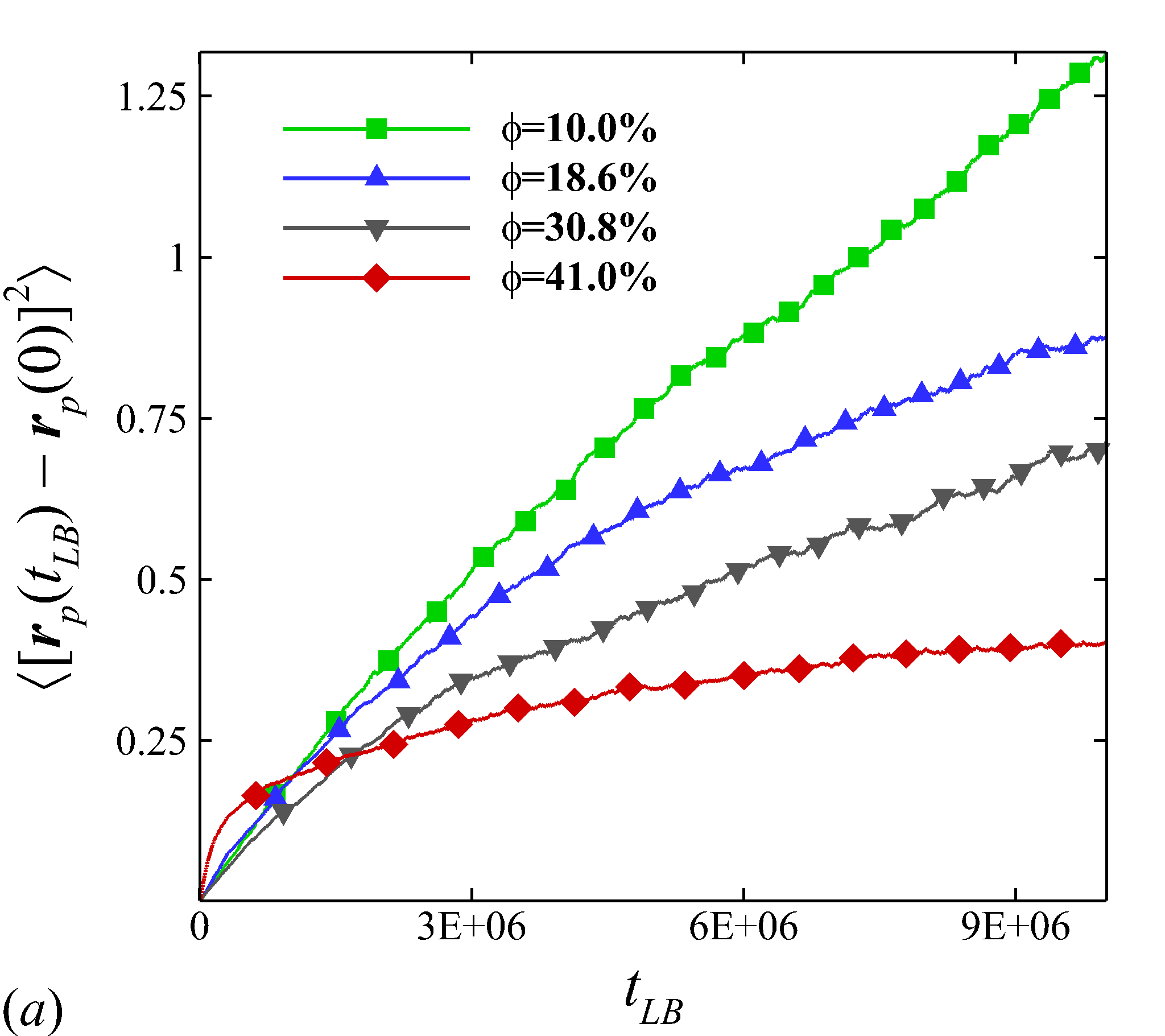}
\includegraphics[width=0.5\columnwidth]{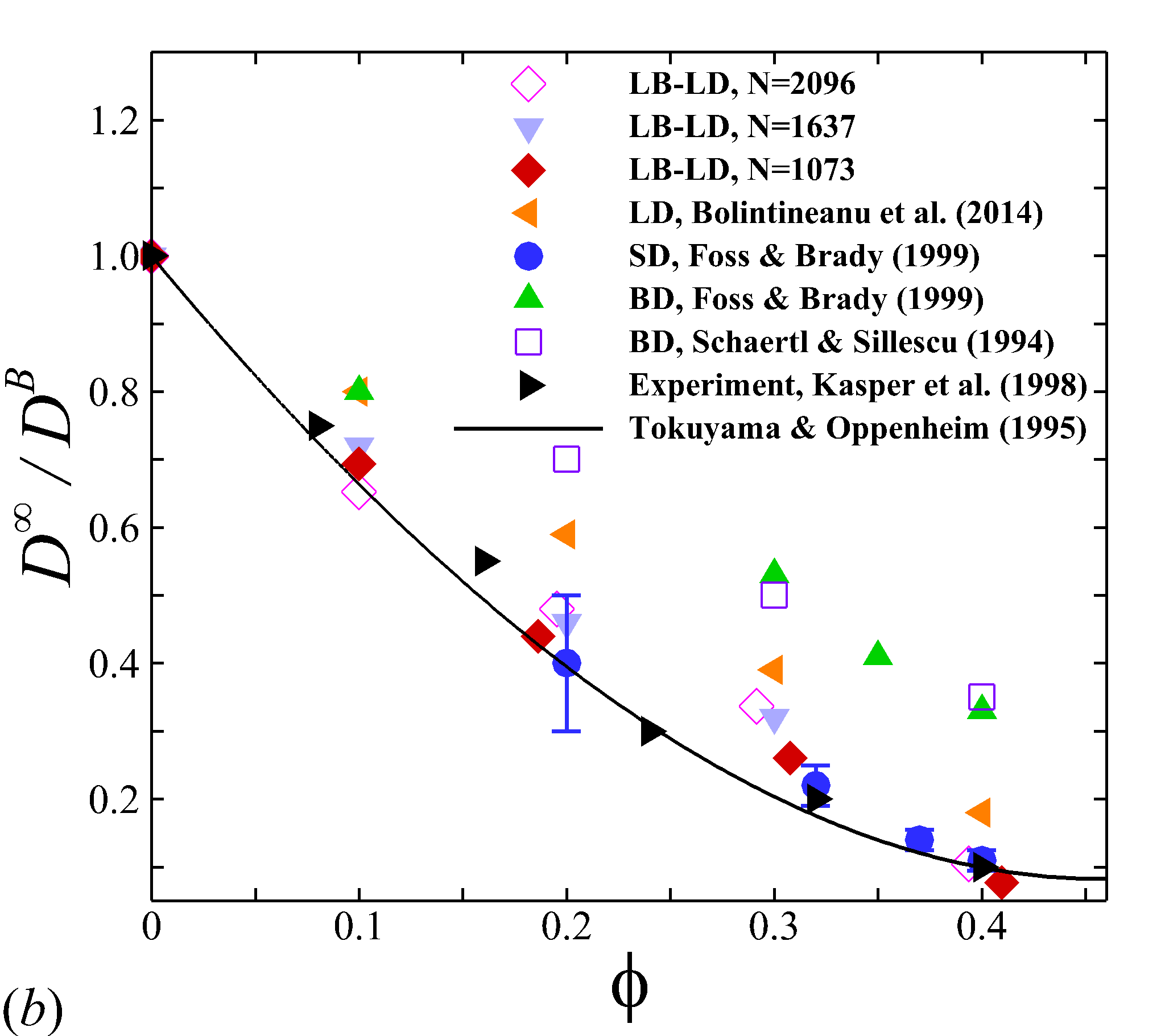}
}
\caption{(a) The mean squared displacements at various particle packing fractions with $N=1073$. (b) The normalized self-diffusivity of colloidal particles as a function of particle volume concentration. Experimental results, theory and particle-scale simulation results are also plotted for comparison purpose.}
\label{fig:diff_c}
\end{figure}

Figure \ref{fig:diff_c} (a) shows the evolution of the particle MSD for various particle packing fractions with $N=1073$. A clear decrease of the MSD slope with increasing particle concentration is observed. Figure \ref{fig:diff_c} (b) further presents the normalized long-time diffusivity against the particle packing fraction. Results from existing simulation studies \citep{FossJFM1999,Schaertl1994,FossJR2000,Bolintineanu2014}, experimental measurements \citep{Kasper1998} and theory \citep{Tokuyama1994} are also plotted for comparison. At zero particle packing fraction, all results agree excellently with the theoretical Brownian diffusivity. At non-zero particle packing fraction, the BD method \citep{Schaertl1994,FossJR2000}, which completely neglects the effect of the fluid-solid coupling, shows the largest deviation from both experiment and theory. By improving the contact modeling, the LD model \citep{Bolintineanu2014} exhibits improved accuracy. The SD method \citep{FossJFM1999} shows excellent accuracy since it fully resolves both the near-field and far-field HI. The LB-LD method, which empolys the DLVO contact modeling and includes the fluid-mediated effect through two-way coupling, show good agreement with both the experiment and theory at low ($\phi=0.1$) and high ($\phi=0.4$) particle packing fraction. The small deviation at moderate concentration ($\phi=0.2\sim0.3$) might be related to the exclusion of the lubrication effect, which plays a vital role at semidilute particle suspensions \citep{Cichocki1999}. Overall, the LB-LD method shows satisfactory accuracy given its simplicity in handling the many-body long-range and short-range interactions.

\subsection{Self-globularization of a cohesive polymer chain}\label{sec:relax}
Polymer chains, such as deoxyribonucleic acid (DNA) \citep{larson2005} and von Willebrand factor (vWF) \citep{Katz2006}, tend to form a collapsed globule conformation in a quiescent solvent. The mechanistic drivers for this process primarily include the Brownian motion and the intra-cohesiveness between adjacent monomers. The polymer longest relaxation time (LRT), $\tau_p$, can be used to characterize the rate of the self-globularization process. The LRT is known to increase with the polymer length defined as the number of monomers, $N$. Specifically, under free-draining (FD) condition wherein the many-body HI is neglected, the polymer LRT suggested by \citet{rouse1953}, $\tau_{p,R}$ , scales as
\begin{equation}
\tau_{p,R}\sim O(N^{2.0});
\label{eqn:rouse}
\end{equation}
when including the effect of HI, the polymer LRT, $\tau_{p,Z}$, according to \citet{zimm1956} should scale as
\begin{equation}
\tau_{p,Z}\sim O(N^{1.5}).
\label{eqn:zimm}
\end{equation}
\begin{figure}
\centerline{\includegraphics[width=0.9\columnwidth]{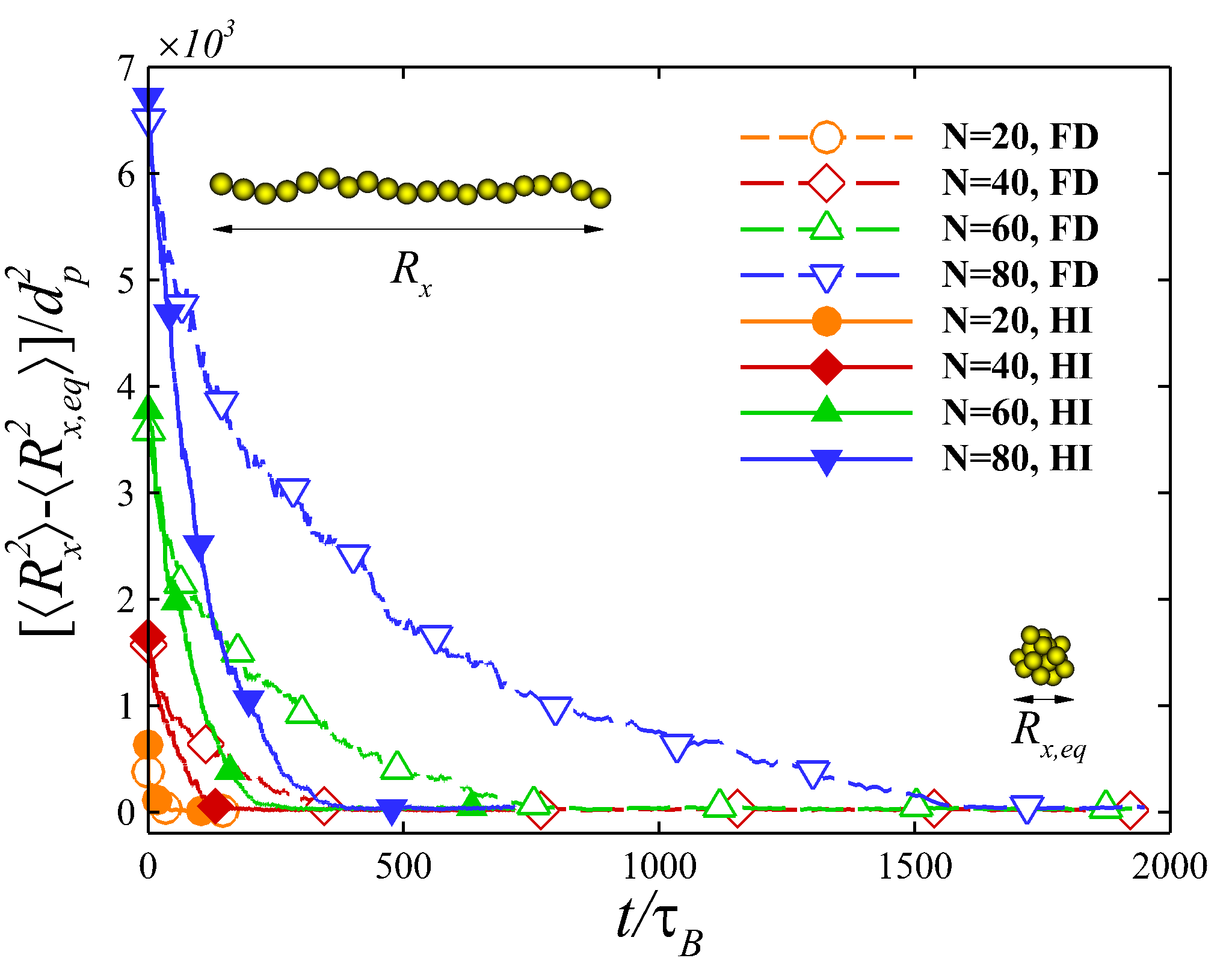}}
\caption{Temporal evolution of normalized MSE, $\langle R_x^2\rangle/d_p^2$, for various polymer length ($N=20\sim80$) under either HI or FD condition. The left inset shows the streched state of the polymer chain, while the right inset shows the globular state (after relaxation) of the polymer chain.}
\label{fig:glb1}
\end{figure}
\begin{figure}
\centerline{
\includegraphics[width=0.5\columnwidth]{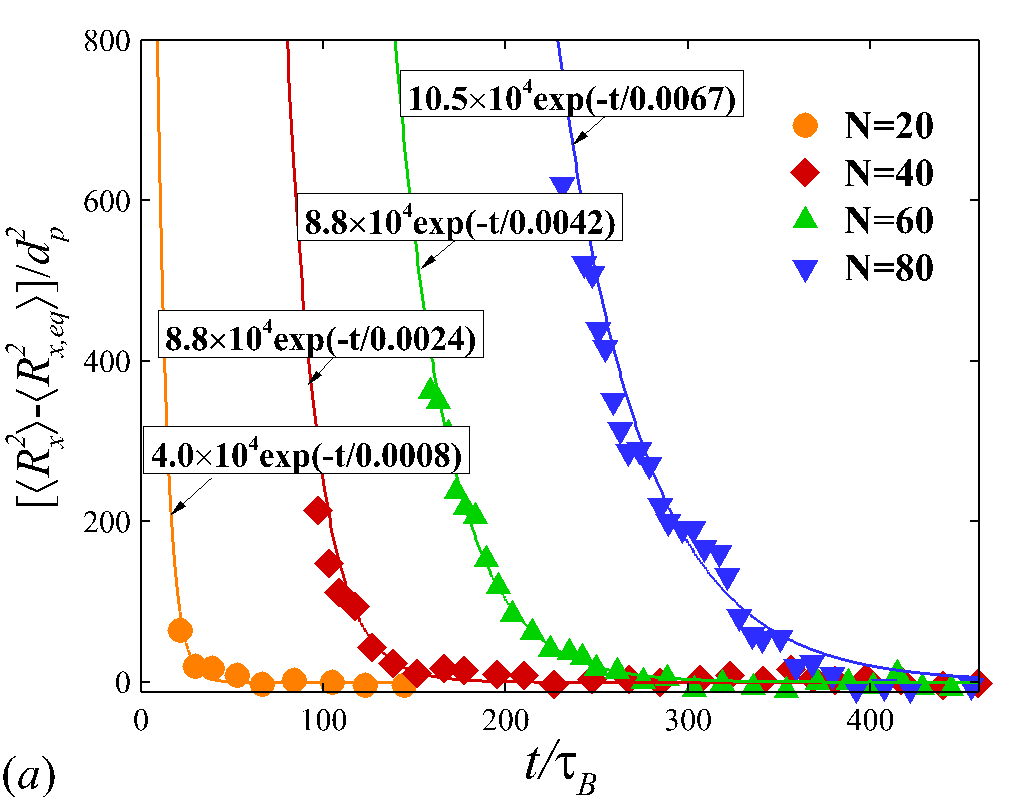}
\includegraphics[width=0.5\columnwidth]{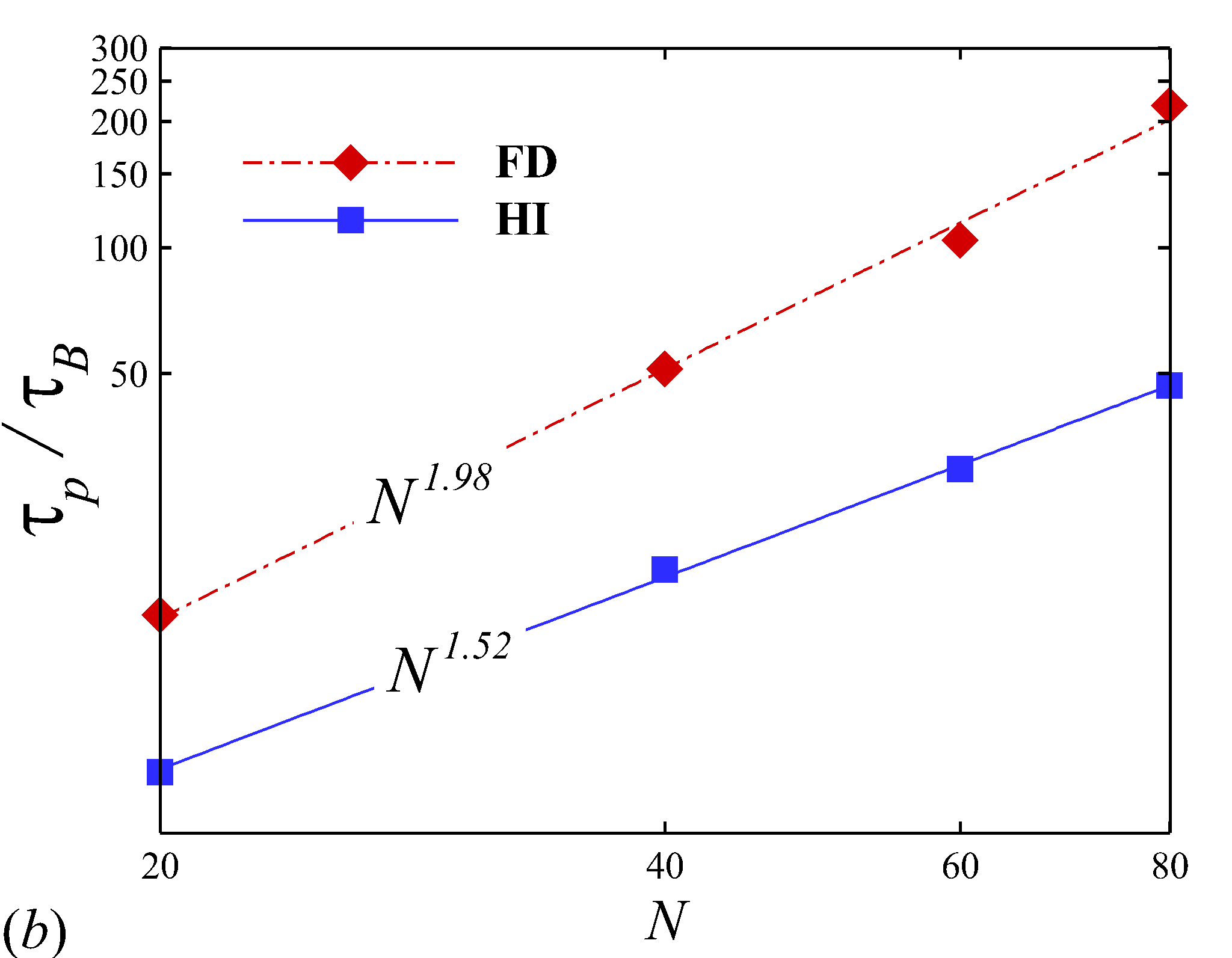}
}
\caption{(a) The instantaneous shifted MSE data points and the corresponding exponential curve fittings for polymers of various length $N$=20, 40, 60, and 80, under hydrodynamic interaction conditions. Only the data points at $\langle R_x\rangle/(Nd_p)\sim0.3$ are considered for the curve fitting procedure. The curve fitting is performed for all points satisfying $\langle R_x\rangle/(Nd_p)\leq0.3$. (b) Polymer longest relaxation time $\tau_p$ as a function polymer length $N$.}
\label{fig:glb2}
\end{figure}
\indent The self-globularization process of a single polymer chain in a quiescent flow under FD or HI conditions is interrogated using the current LB-LD approach. Four polymer lengths, $N$=20, 40, 60 and 80, are considered to obtain a scaling observation. The diameter of each monomer is set to $d_p$=60 $nm$. \textcolor{black}{A periodic computational domain with dimensions of $100^3$ is selected to minimize the periodic boundary effect.} The polymer extension, $R_x$, defined as the polymer projection length along the $\text{X}$ direction (streamwise direction), is introduced to quantify the instantaneous conformation of the polymer. The over-damped LE is employed, given the time scale of the polymer conformational change is much larger than the particle relaxation time scale. Figure \ref{fig:glb1} plots the normalized mean-square polymer extension (MSE), $\langle R_x^2\rangle/d_p^2$, versus normalized time, $t/\tau_B$. Each curve is obtained through ensemble average over three independent runs with the same polymer length. The insets of Figure \ref{fig:glb1} demonstrate the initial and final conformation of a polymer chain with $N=20$. For both the FD and HI cases, longer polymers (larger $N$) take longer time to reach equilibrium globular conformation. Additionally, polymers of the same length under HI conditions tend to tend to form a globule conformation more rapidly compared to under FD conditions. The underlying mechanism causing this time-scale discrepancy between HI and FD conditions is that the flow pattern induced by the polymer itself leads to smaller relative velocity and hence smaller drag force exerted on each monomer \citep{piotr2011}.

The polymer relaxation time $\tau_p$ can be quantitatively evaluated by fitting the instantaneous MSE, $[\langle R_x^2\rangle-\langle R_{x,eq}^2\rangle]/d_p^2$, with an exponential function form in the final relaxation regime when $\langle R_x\rangle/Nd_p\leq0.3$, as suggested in \citet{Perkins1997}. The obtained exponential curve satisfies the following expression
\begin{equation}
\frac{\langle R_x^2\rangle-\langle R_{x,eq}^2\rangle}{d_p^2}=\mathcal{A}e^{-t/\tau_p},
\label{eqn:relx1}
\end{equation}
where $\mathcal{A}$ is a fitting constant, and the denominator of the exponent readily gives $\tau_p$. Figure \ref{fig:glb2} (a) presents the exponential curve fittings for the instantaneous $[\langle R_x^2 \rangle-\langle R_{x,eq}^2\rangle]/d_p^2$ data points at $\langle R_x\rangle/Nd_p\leq0.3$ under HI condition. The corresponding exponential functions are denoted beside each curve as shown in Figure \ref{fig:glb2} (a). Same curve fitting procedure is also performed for the FD case to obtain the LRT of the polymer chain under FD condition, which is not shown here for clarity. The obtained $\tau_p$ for both FD and HI cases are further plotted against the polymer length $N$ in Figure \ref{fig:glb2} (b), where the two straight lines are the best power-law fits for $\tau_{p,R}$ and $\tau_{p,Z}$ data points, respectively. The relaxation time $\tau_p$ of the single polymer chain captured based on the LB-LD approach show scaling exponents of 1.98 and 1.52 for the FD and the HI conditions, respectively. These scaling behaviors agree nicely with the theoretical scaling arguments \citep{rouse1953,zimm1956}. These results show the conformational dynamics of a single polymer chain subject to many-body HI is favorably captured through the current LB-LD approach. It also confirms the significance of including the effect of HI for simulating nanoscale particulate suspensions.

\subsection{Shear-induced unfolding of a collapsed polymer chain}\label{sec:vWF}
\begin{figure}
\centerline{\includegraphics[width=1.0\columnwidth]{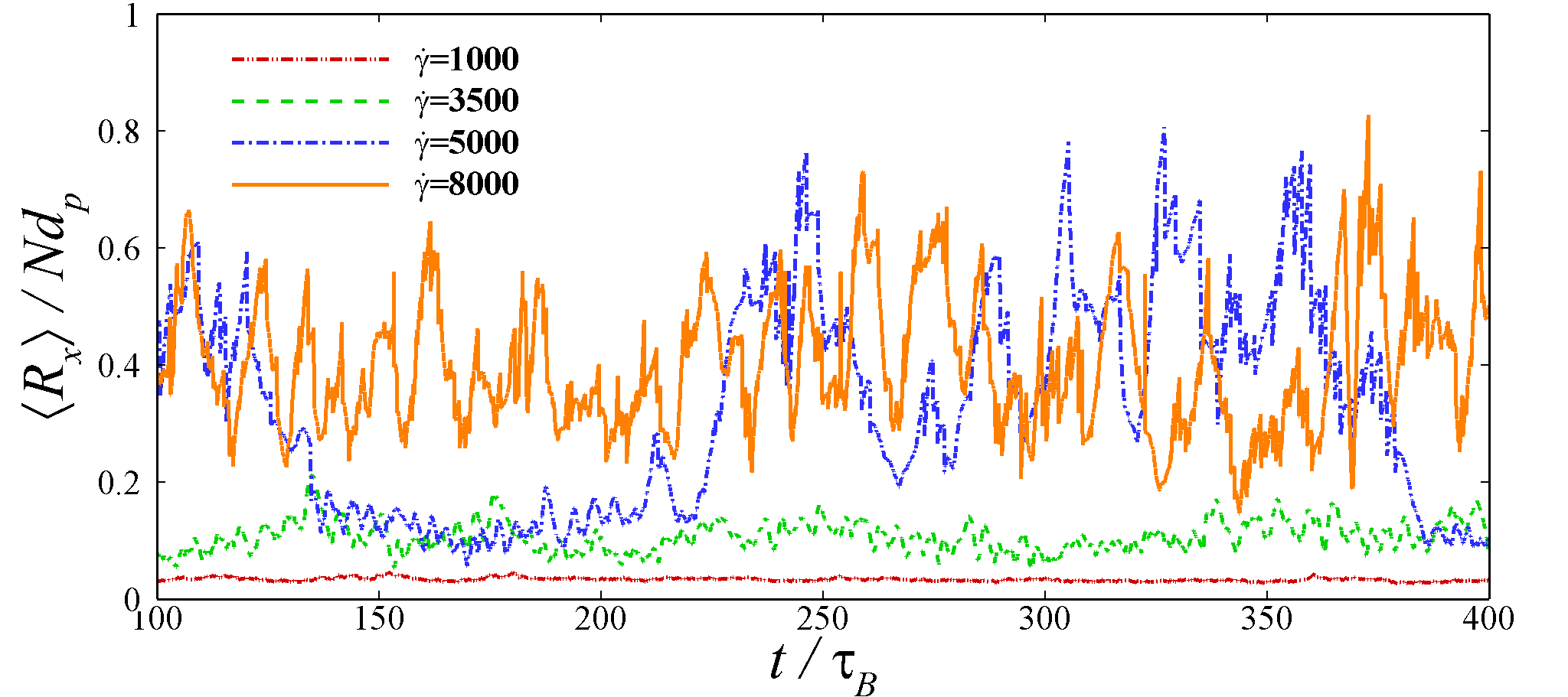}}
\caption{The temporal evolution of the normalized mean extension of a single polymer chain (model vWF) subject to various shear rates in an unbounded simple shear flow.}
\label{fig:evl}
\end{figure}
When subject to shear flow above certain critical shear rate, the collapsed biopolymer chains (e.g. vWF, DNA, etc.) tend to be unfolded and exhibit periodic stretch/coil conformation changes while tumbling \citep{larson2005,Katz2006,schneider2007shear}. Such shear-induced unfolding phenomenon of the collapsed polymer chain is a manifestation of the competition between the viscous, elongational flow effect and the intra-monomer cohesive effect \citep{Katz2006}. In this section, we further validate the LB-LD approach by simulating a single vWF strand subject to different shear rates and comparing its conformational statistics with existing experimental data. To exclude the confinement effect, the Lees-Edwards boundary condition \citep{LEbc1972,ClausenJFM2011} is employed to impose the unbounded simple shear flow. Shear rates, ranging from 1 to 8\ 000 $s^{-1}$, are considered to replicate the flow conditions considered in the experiment \citep{schneider2007shear}. The vWF strand is modeled as a 200-bead (i.e. $N=200$) polymer chain. Each bead represents a dimer (repeating unit of a vWF polymer chain) with a diameter of $d_p=160$ $nm$ closely matching measured dimer size reported in \citet{Springer2016}. The selected bead size and bead number yield a contour length of $32\ \mu m$ for the model vWF, which is also close to the actual size of the vWF considered in the experiment \citep{schneider2007shear}. \textcolor{black}{All computations adopt a periodic LB domain with dimensions of 60$\times$30$\times$30 $\mu m^3$ in the flow, the velocity-gradient and the vorticity directions, respectively.}

\begin{figure}
\centerline{\includegraphics[width=0.9\columnwidth]{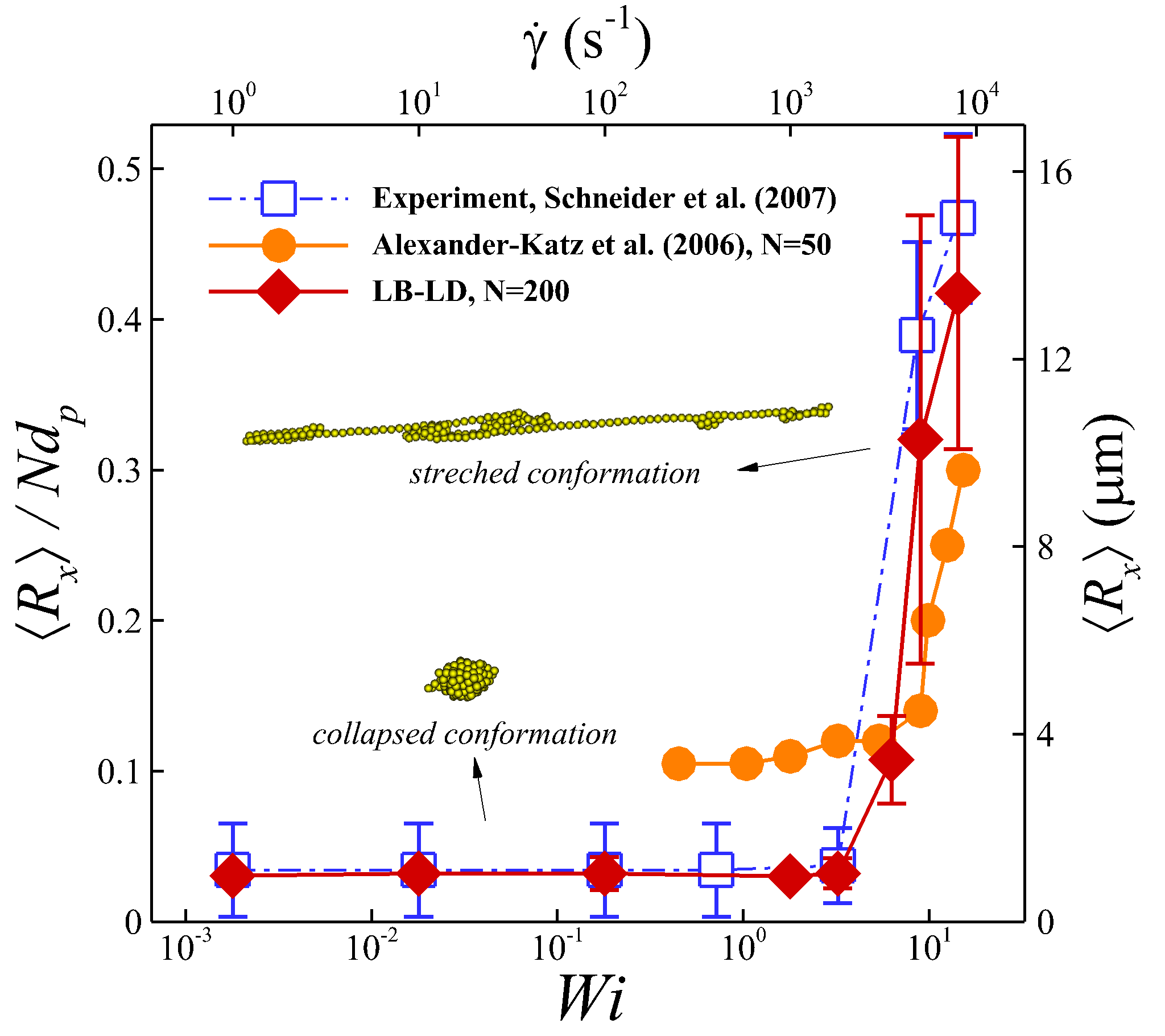}}
\caption{Unfolding of a single polymer chain under a critical shear rate. The measured polymer mean extension using a microfluidic device by \citet{schneider2007shear} is adopted to validate the LB-LD model. The simulation results reported in \citet{Katz2006} are also plotted for comparison. The right vertical axis (dimensional) applies to experimental results and the LB-LD results; the left vertical (dimensionless) axis applies to all three sets of results. }
\label{fig:unfolding}
\end{figure}

The conformation state of a single polymer chain under shear flow can be quantified by the normalized mean polymer extension (ME) in the streamwise direction, $\langle R_{x}\rangle/Nd_p$, where the ensemble average of the polymer extension is performed in time. As shown in Figure \ref{fig:evl}, the instantaneous polymer extension exhibits fluctuation over time. Moreover, the magnitude of the fluctuation and the ME increase with shear rate changing from 1\ 000 $s^{-1}$ to 8\ 000 $s^{-1}$. Figure \ref{fig:unfolding} plots the ME of the polymer chain versus shear rate in both dimensional and dimensionless units. A Weissenberg number, defined as $Wi=\dot{\gamma}\tau_{B}$, is employed as the nondimensional shear rate. It is shown that the ME measured in the LB-LD simulation compares favorably well with the experimental measurements reported in \citet{schneider2007shear}. Particularly, a range of critical shear rates (5\ 000 $\sim$6\ 000 $s^{-1}$) around which the polymer chain exhibits an abrupt increase of the ME is well captured in the LB-LD simulation. The insets in Figure \ref{fig:unfolding} presents the representative polymer conformations under simple shear flow. As expected, below critical shear rate, the vWF polymer remains in a compact globular conformation; while above critical shear rate, the vWF polymer undergoes periodically stretched/coiled conformational transitions. The simulation results reported by \citet{Katz2006} using Brownian dynamics are also adopted for comparison, where the critical shear rate is shown to be well captured. However, their simulation deviates from the experimental results in terms of normalized mean extension particularly at low shear rates, which is due to the shorter contour length ($N=50$) considered in their model vWF. Since the vWF model parameters (except $N$) considered in the LB-LD polymer model is close to the ones employed in \citet{Katz2006}, the results also qualitatively confirm the weak polymer length dependence of the critical shear rate as concluded in \citet{Katz2006}. The good agreement between the LB-LD simulation and experiment results for the shear-induced unfolding process of vWF strand again show the validity of the current LB-LD approach in capturing the conformational dynamics of long-chain polymers under shear flow.

\section{Summary and conclusions}\label{sec:con}
A hybrid Eulerian-Lagrangian approach coupling the non-fluctuating LB method and a LD method is developed to simulate suspensions of nanoscale particles and long-chain polymers including the effects of thermal fluctuation, many-body HI, and particle-particle short-distance interactions with linear particle-number scalability. The LB-LD approach is verified and validated with both thoery and experiment. An EH algorithm is also developed to handle the short-range pairwise particle search and interaction, which ensures localization and hence linear scalability of the method while performing particle neighbour search. The LB-LD approach embedded with the EH algorithm is purely local and can be readily extended for parallelization. 

The LB-LD approach confirms the idea of coupling non-fluctuating LB method with LD method to directly capture the correct Brownian diffusion without empirical rescaling of the particle mobility \cite{mynam2011}. Furthermore, the current method demonstrates that the long-distance many-body HI can be directly included via the LB-LD two-way coupling scheme, which was not shown in \citet{mynam2011}. Compared to using mobility matrix approach to capture HI \cite{Katz2006}, the direct two-way coupling approach, in addition to being more efficient, also has the flexibility of including the modified HI effects subject to complex geometries/boundaries \cite{Liu2018a,Liu2018JFM}. Besides, the two-way coupled LB-LD approach embedded with the DLVO potentials allows simulating nanoscale particulate suspension across dilute-to-dense concentrations with good accuracy.

\begin{figure}
	\centerline{
		\includegraphics[width=0.705\columnwidth]{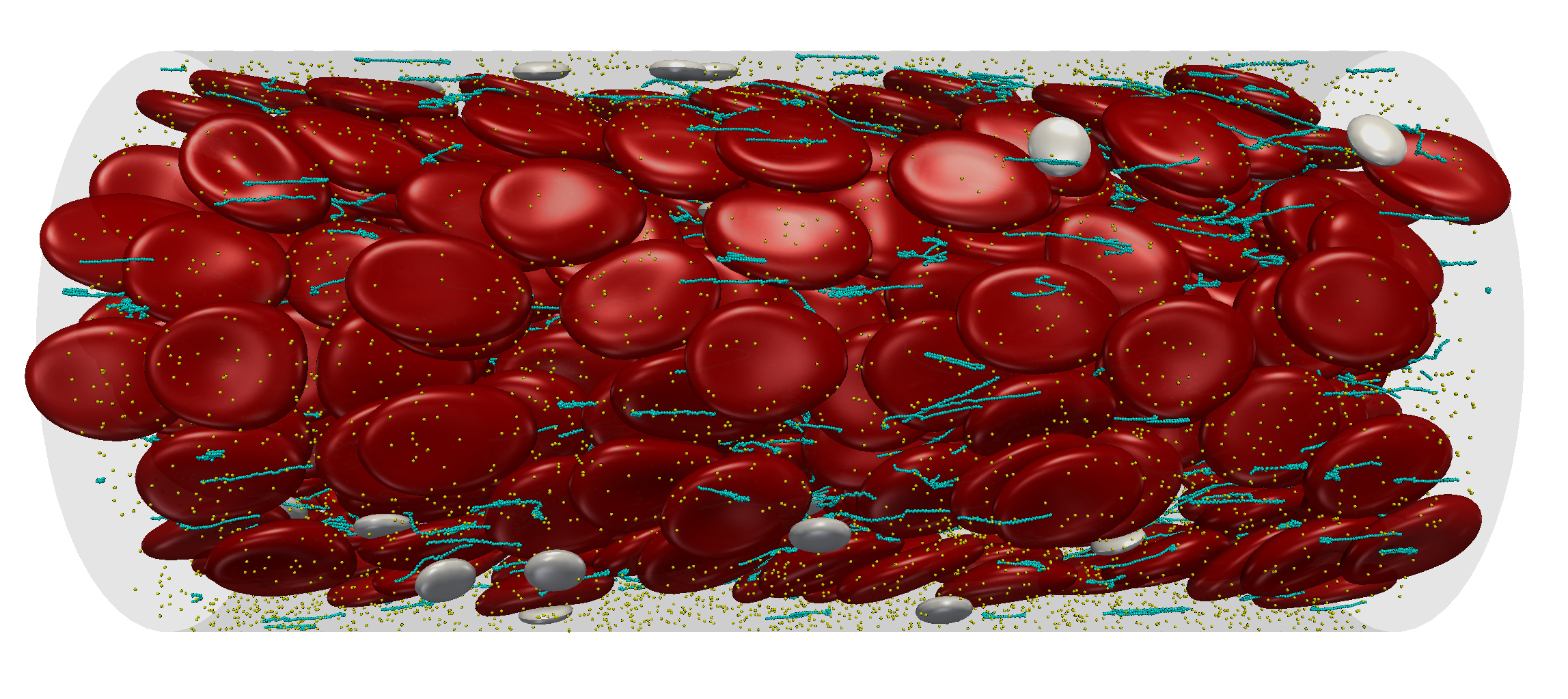}
		\includegraphics[width=0.295\columnwidth]{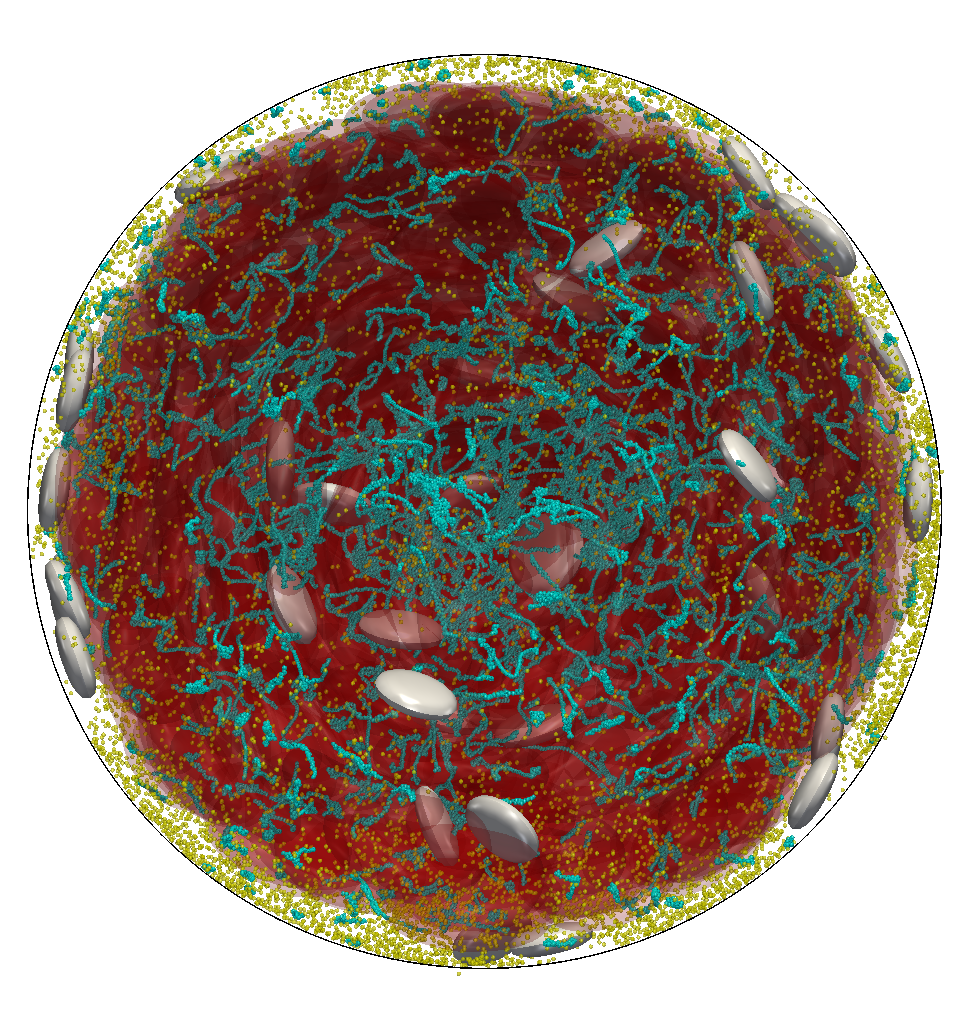}
	}
	\caption{The side view (left) and frontal view (right) of complex whole blood flow through a microvessel with a 40 $\mu m$ diameter, simulated using the LB-LD method coupled with the spectrin-link method \citep{Reasor2012}. A total of 427 deformable red blood cells (RBCs) (red capsules) are resolved in the simulation, resulting in a 40\% heamatocrit. The platelet (white oblate capsules)-RBC number ratio is $\sim 1/20$. The vWF (light blue strands) molecule number concentration is $\sim 3\times10^{10}/ml$. The wall shear stress is $2000\ s^{-1}$. All above parameters match the blood physiological condition. The nanoparticle (yellow particles) number concentration is $\sim 1\times10^{9}/ml$ consistent with the typical dosage used in human.}
	\label{fig:all}
\end{figure}

Since the entire nanoscale particulate suspension dynamics are resolved through sub-lattice techniques, the LB-LD method is particularly suitable for multimodal particulate suspension applications that involve particles, polymer chains and capsules with disparate length scales \citep{Liu2018a,Liu2018JFM}, where DNS of such systems is computationally prohibitive.
\textcolor{black}{One example is the simulation of multiscale, multicomponent complex blood flow by coupling the LB-LD method with cellular blood solvers, where the dynamics and deformation of both microscale blood cells (e.g. red blood cell, platelet, etc.) and nanoscale molecules and bioproteins (e.g., vWF, albumen, etc) can be simulated concurrently without neither refining the grids nor introducing sub-timesteps. In Figure \ref{fig:all}, we present a simulation snapshot demonstrating a simulation of multiscale and multicomponent whole blood through a 40 $\mu m$ microvessel. This simulation captures a concentrated poly-dispersed suspension tubular flow of 40\% volume fraction of deformable red blood cells (red), $\sim$1\% volume fraction of rigid platelets, 3$\times$$10^{10}$/$ml$ vWF and 1$\times$$10^9$/$ml$ nanoscale particles under a wall shear rate of 2000 $s^{-1}$. The development of the LB-LD method shows promise in forming a multiscale computational framework to tackle biophysical suspension flow problems across nano-to-microscale, such as high-shear induced thrombus formation in blood \cite{Casa2017}.}


\section*{Acknowledgements}
The authors acknowledge the financial support from Sandia National Laboratories under grant number 2506X36 and the computational resource granted by the Extreme Science and Engineering Discovery Environment (XSEDE) of National Science Foundation under grant number TG-CT100012. Sandia National Laboratories is a multimission laboratory managed and operated by National Technology and Engineering Solutions of Sandia LLC, a wholly owned subsidiary of Honeywell International Inc. for the U.S. Department of Energy’s National Nuclear Security Administration under contract DE-NA0003525. This paper describes objective technical results and analysis. Any subjective views or opinions that might be expressed in the paper do not necessarily represent the views of the U.S. Department of Energy or the United States Government.



\bibliographystyle{model1-num-names}
\bibliography{LBLD}

\end{document}